\documentclass[bookmarks=true,pdfnewwindow=true,
colorlinks=true,linkcolor=xlinkcolor,citecolor=xlinkcolor,filecolor=xlinkcolor,
urlcolor=xlinkcolor,final=true,allcolors=blue]{aastex701}

\usepackage[english]{babel}

\usepackage{array}
\usepackage{graphicx}
\usepackage{comment}

\usepackage{algorithm}
\usepackage{algpseudocode}

\usepackage[normalem]{ulem}

\usepackage{amsmath}
\usepackage{amsthm}
\usepackage{amssymb}
\usepackage{mathrsfs}

\usepackage{multirow}
\usepackage{booktabs}

\newcommand{\rone}{r_{\nu x}}
\newcommand{\ronex}{r_{x\nu}}
\newcommand{\rtwo}{r_{z\theta}}
\newcommand{\rthree}{r_{\theta\phi}}

\usepackage{xcolor}

\newif\ifshowchanges
\showchangesfalse

\ifshowchanges
  \newcommand{\revtext}[1]{\textcolor{red}{#1}}
  \newenvironment{revblock}{\color{red}}{}
\else
  \newcommand{\revtext}[1]{#1}
  \newenvironment{revblock}{}{}
\fi

\begin{document}

\title{Multigroup Thermal Radiation Transport with Tensor Trains}

\author[0009-0009-8285-861X]{Aditya S. Deshpande}
\affiliation{Department of Aerospace Engineering, University of Michigan, Ann Arbor, MI}
\affiliation{Michigan SPARC, Los Alamos National Laboratory, Ann Arbor, MI}
\email{dadity@umich.edu}

\author[0000-0003-2131-4634]{Patrick D. Mullen}
\affiliation{Michigan SPARC, Los Alamos National Laboratory, Ann Arbor, MI}
\affiliation{Computing and Artificial Intelligence, Los Alamos National Laboratory, Los Alamos, NM}
\email{pdmullen@lanl.gov}

\author[0000-0003-3152-8206]{Alex A. Gorodetsky}
\affiliation{Department of Aerospace Engineering, University of Michigan, Ann Arbor, MI}
\affiliation{Michigan SPARC, Los Alamos National Laboratory, Ann Arbor, MI}
\email{goroda@umich.edu}

\author[0000-0003-4353-8751]{Joshua C. Dolence}
\affiliation{Michigan SPARC, Los Alamos National Laboratory, Ann Arbor, MI}
\affiliation{Computing and Artificial Intelligence, Los Alamos National Laboratory, Los Alamos, NM}
\affiliation{Department of Aerospace Engineering, University of Michigan, Ann Arbor, MI}
\email{jdolence@lanl.gov}

\author[0000-0002-7530-6173]{Chad D. Meyer}
\affiliation{Michigan SPARC, Los Alamos National Laboratory, Ann Arbor, MI}
\affiliation{Continuum Models and Numerical Methods, Los Alamos National Laboratory, Los Alamos, NM}
\email{chadmeyer@lanl.gov}

\author[0000-0001-6432-7860]{Jonah M. Miller}
\affiliation{Michigan SPARC, Los Alamos National Laboratory, Ann Arbor, MI}
\affiliation{Computing and Artificial Intelligence, Los Alamos National Laboratory, Los Alamos, NM}
\email{jonahm@lanl.gov}

\author[0000-0001-7364-7946]{Luke F. Roberts}
\affiliation{Michigan SPARC, Los Alamos National Laboratory, Ann Arbor, MI}
\affiliation{Computing and Artificial Intelligence, Los Alamos National Laboratory, Los Alamos, NM}
\email{lfroberts@lanl.gov}

\begin{abstract}
    We investigate the application of tensor-train (TT) algorithms to multigroup thermal radiation transport (i.e., photon radiation transport). The TT framework enables simulations at discretizations that might otherwise be computationally infeasible on conventional hardware. We show that solutions to certain multigroup problems possess an intrinsic low-rank structure, which the TT representation leverages effectively. This enables us to solve problems where the discretized solution size exceeds a trillion parameters on a single node. The solver is evaluated on a range of test problems with varying levels of complexity, consistently achieving compression factors greater than $100 \times$ and speedups exceeding $2 \times$. We also investigate alternative TT topologies by analyzing the low-rank structure of the merged spatio-spectral core to assess the potential for greater compression. This analysis suggests that compression gains could increase by factors as large as $7$. Our results indicate that the low-rank structure of the merged spatio-spectral core captures the spatio-spectral complexity of the solution, largely driven by the opacity structure of the medium. Beyond identifying opportunities for improved compression, this analysis highlights the types of errors that may arise in angle-integrated quantities when exploiting this low-rank structure.
\end{abstract}


\section{Introduction}
Frequency-dependent thermal radiation transport is central to a wide range of applications in astrophysics—including line-driven winds \citep[e.g.,][]{shlosman85, proga2000}, black hole accretion \citep[see, e.g.,][and references therein]{zhang25a}, supernova explosions \citep[e.g.,][]{chevalier08, wollaeger2013radiation, goldberg22}, star formation \citep[e.g.,][]{gonzalez15, rosen2016unstable, rosen20, menon2023outflows}, gamma-ray bursts \citep[e.g.,][]{thompson1994model, zrake2019subphotospheric, nedora2025multiphysics}, tidal disruption events \citep[e.g.,][]{huang25, giron26}, and X-ray pulsars \citep[see,][and references therein]{caballero2012xraypulsarsreview}---as well as in nuclear science and engineering, particularly inertial confinement fusion \citep[ICF;][]{haines22, kim23}.  

The radiation transport equation for photons governs the specific intensity $I_\nu$, a quantity defined over space, propagation direction, and frequency. This intrinsic high dimensionality renders first-principles numerical solutions extremely expensive, motivating a hierarchy of approximations that reduce angular and/or spectral resolution. One common simplification is the gray approximation, in which frequency dependence is removed by integrating the intensity over the entire spectrum. While this eliminates one dimension of the problem, it implicitly assumes that frequency-dependent opacity effects can be well represented by suitable averages. Multigroup radiation transport (MRT) relaxes this assumption by discretizing the frequency domain into a finite set of groups, retaining limited spectral resolution while remaining computationally tractable.
However, even within the multigroup framework, solving the transport equation directly remains a challenge. Discrete ordinates $S_N$ methods require resolving the angular domain with many directions \citep[to eliminate so-called ``ray effects"; see, e.g.,][]{gorodetsky25} and the spectral domain with multiple frequency groups, leading to a rapid growth in the total number of required parameters to store the solution. This burden is further amplified by the need for fine spatial discretizations in many applications---for example, to accurately resolve material interfaces, shocks, or fluid instabilities---thereby exacerbating the curse of dimensionality. 

This curse of dimensionality is particularly severe in applications such as line-driven transport and indirect-drive inertial confinement fusion (ICF).  In stellar and active galactic nuclei (AGN) winds, the line-driving force arises from interactions with many ultraviolet resonance lines \citep{proga2007theory}. The cumulative radiative force depends sensitively on how the radiation field samples a dense forest of line opacities as a function of frequency. Accurately representing this physics directly in the transport equation requires a finely resolved frequency grid to capture overlapping lines and ionization-dependent features, often necessitating hundreds to thousands of frequency groups in the UV alone.  In ICF hohlraums, X-ray drive and wall re-emission involve complex, strongly frequency-dependent opacities with sharp edges and line features associated with high-Z materials \citep{tolkach2003development}. Capturing non-gray effects in absorption, re-emission, spectral shaping, drive symmetry, and capsule preheat requires MRT with tens to hundreds of energy groups.

To make MRT problems computationally tractable, existing $S_N$ calculations typically have been limited to relatively few groups, few angles, short runtimes, and/or limited spatial resolutions \citep[e.g.,][]{secunda24, mills24, secunda25, jiang25, liu25, huang25}.  More commonplace, an alternative approach to mitigating the inherent dimensionality of the transport equation is to evolve angle-integrated quantities using moment-based methods. Such approaches have been extended to the multigroup setting, including flux-limited diffusion FLD and M1 models \citep[see, e.g.,][for a nice review]{wunsch24}. However, like their gray counterparts, these methods suffer from limitations associated with angular averaging. In addition to invoking moment-based methods, contemporary ICF simulations also frequently reduce dimensionality by restricting the spatial geometry to one- or two-dimensional settings. In the context of line-driven winds,  \cite{proga99} replaces full transport with so-called ``force multiplier" models.  

In this work, we propose to address the curse of dimensionality by incorporating tensor-trains \citep[TT,][]{oseledets2011tensor}  into a multigroup transport solver. Tensor-train representations have been successfully applied to gray thermal transport \citep{gorodetsky25}, and were able to reproduce analytical solutions across numerous test cases while achieving speedups and compression factors of $\sim$60$\times$ and $\sim$1000$\times$, respectively. This naturally raises the question of whether solutions to the MRT equations also exhibit low-rank structure and can similarly be represented efficiently using TTs. We explore this question by developing a multigroup TT solver. Additionally, given the many possible tensor-network representations of the solution, we also investigate alternative TT topologies and compare their performance. The remainder of the paper is structured as follows. \S \ref{sec:governing_eqns} presents the MRT equations and outlines the numerical algorithm, while \S \ref{sec:tensors} details the tensorization of the solver. We conduct a series of multigroup test problems to assess the solver in \S \ref{sec:tests} followed by an extensive discussion in \S \ref{sec:discussion}. \S \ref{sec:conclusion} concludes the work with a summary of our findings.

\section{Governing equations} \label{sec:governing_eqns}
The time evolution of the frequency-dependent specific intensity $I_\nu$ is governed by the transport equation \citep{1984frh..book.....M}
\begin{equation}
    \partial_t I_\nu + c \mathbf{n} \cdot \nabla I_\nu = c \left(j_\nu - \alpha_\nu I_\nu \right),
    \label{eq:generalRT}
\end{equation}
where $c$ is the speed of light, $\mathbf{n}$ is a unit normal defining photon propagation direction, and $j_\nu$ and $\alpha_\nu$ denote the frequency-dependent emissivity and absorptivity, respectively. 

In the multigroup formalism, the frequency domain is divided into $N_\nu$ bins. End points $\nu_0, \nu_1, \dots, \nu_{N_\nu}$ with $\nu_0 = 0$ and $\nu_{N_\nu}=\infty$ cover the entirety of frequency space. Integrating the transport equation over a frequency bin $f$ with frequency bounds $[\nu_{f-1}, \nu_{f})$ results in
\begin{equation}
    \partial_t I_f + c \mathbf{n} \cdot \nabla I_f = c \left(j_f - \alpha_f I_f \right),
    \label{eq:MGRT}
\end{equation}
where the $f$ subscript denotes \revtext{a group over which some quantity is integrated} (e.g., $I_f = \int_{\nu_{f-1}}^{\nu_f} I_\nu \mkern3mu d\nu$ is the frequency-integrated specific intensity in bin $f$) and the group integrated absorptivity $\alpha_f$ is assumed to be an average over the group.\footnote{Planck or Rosseland averaging could be employed in constructing group mean opacities.  For simplicity, we do not augment the transport equation to include \textit{both} Planck and Rosseland opacities as done in \cite{jiang22}, which would otherwise amount to an additional contribution to the coupling source term.}  

Assuming a static background medium (with material mass density $\rho$ and temperature $T$), local thermodynamic equilibrium (LTE), and isotropic, elastic scattering, we can expand the multigroup transport equation as
\begin{equation}
    \partial_t I_f + c \textbf{n} \cdot \nabla I_f = c [ \rho \kappa_{s,f} (J_f - I_f) + \rho \kappa_{a,f} (\varepsilon_f - I_f )],
    \label{eq:finalMGRT_assumptions}
\end{equation}
where $\kappa_{a,f}$ and $\kappa_{s,f}$ are the specific group mean absorption and scattering opacities, respectively, $J_f$ is the group mean intensity over solid angle $\Omega$ (related to the group radiation energy density via $E_f = \frac{4 \pi}{c} J_f = \frac{1}{c} \int I_f \: d\Omega$), and $\varepsilon_f$ is an emission coefficient proportional to the integral of the Planck function $B(\nu, T)$ over $[\nu_{f-1}, \nu_{f})$
\begin{align}
    \varepsilon_f &= \frac{c}{4 \pi} \int_{\nu_{f-1}}^{\nu_{f}} B(\nu, T) d\nu,\label{eq:Ttoeps1} \;\;\ \mathrm{where}\\
    B(\nu, T) &= \frac{8 \pi h \nu^3}{c^3} \frac{1}{\exp(h \nu/\left[k T\right]) - 1}.\label{eq:Ttoeps2}
\end{align}
For a single group  with $[\nu_{0}, \nu_{N_\nu}) \rightarrow [0,\infty)$ (i.e., ``gray"), 
$\varepsilon_f$ reduces to $\varepsilon_\mathrm{gray} = c a T^4 / [4 \pi]$, where $a$ is the radiation constant.  

Radiation feedback on a fluid is determined via conservation of total energy hence giving rise to a material temperature equation
\begin{equation}
    \partial_t T = -\sum_{f} \left[ \frac{c \kappa_{a,f}}{c_v} \left(\frac{4 \pi}{c} \varepsilon_f - E_f \right) \right],
    \label{eq:Tempeqn}
\end{equation}
\revtext{where $c_v$ denotes the specific heat capacity (at constant volume) of the material.}
    
\subsection{Numerics} \label{sec:numerics}
In this work, the \cite{gorodetsky25} thermal radiation transport with tensor trains (TTTT) solver is extended to multigroup. We highlight here only the modifications to the numerical algorithm that pertain specifically to transitions between the gray and multigroup formalisms.

The transport operator is evolved \revtext{via forward Euler} using the same three flux stencils from \cite{gorodetsky25}: upwinding, Rusanov \citep{toro2009}, and the asymptotic-preserving, HLL flux of \cite{jiang21}.  The introduction of multigroup changes only the HLL stencil, which now invokes an optical depth per frequency bin $\tau_{c,f}$ (see later, \S\ref{sec:tensors})---wavespeeds are otherwise computed identically. We advance the right hand side of Equation (\ref{eq:finalMGRT_assumptions}) from time index $m$ to $m+1$ via a backward Euler, operator split update,
\begin{align}
    \frac{I_f^{m+1} - I_f^m}{c\Delta t} &= \rho^m \kappa^m_{s,f} (J_f^{m+1} - I_f^{m+1}) + \rho^m \kappa^m_{a,f}(\varepsilon_f^{m+1} - I_f^{m+1}),
    \label{eq:discRTsource} \\
    \frac{T^{m+1} - T^{m}}{c\Delta t} &= -\sum_{f} \left[ \frac{\kappa^m_{a,f}}{c^m_v} \frac{4\pi}{c}\left(\varepsilon_f^{m+1} - J_f^{m+1} \right) \right],
    \label{eq:Tupdate}
\end{align}
where $\kappa_{a,f}$ and $\kappa_{s,f}$ are assumed to be constant over the coupling stage.  Integrating Equation (\ref{eq:discRTsource}) over solid angle gives an \revtext{expression for} the advanced mean intensity \revtext{$J_f^{m+1}$} that can be substituted into Equation (\ref{eq:Tupdate}) to arrive at a \revtext{function}
\begin{equation}
    f(T^\dagger) \equiv T^m - T^\dagger - \sum_f \left[
            \frac{4 \pi \Delta t \;\kappa^m_{a,f}}{c^m_v}
            \left(
                \varepsilon_f(T^\dagger)
                - \left(\frac{1}{\left[c \Delta t\right]^{-1} + \rho^m \kappa^m_{a,f}}\right) \left(
                    \frac{J_f^m}{c \Delta t}
                    + \rho^m \kappa^m_{a,f} \varepsilon_f(T^\dagger)
                \right)
            \right)
        \right]
\end{equation}
    where the physical root gives an advanced stage temperature $T^\dagger$ (and hence emission coefficient $\varepsilon_f[T^\dagger]$) to implicitly advance the specific intensity $I_f$.  Rather than advancing the temperature via $T^{m+1} = T^\dagger$, we invoke total energy conservation
\begin{equation}
    \rho^m c^m_v (T^{m+1} - T^{m}) = \frac{4 \pi}{c} \sum_f \left[J_f^m -\frac{1}{4\pi} \sum_{\ell,p} I^{m+1}_{f\ell p} \Delta \Omega_{\ell p} \right].
\end{equation}
to govern the feedback of the radiation field on the fluid. 

The integrals of the Planckian used in constructing $\varepsilon_f[T]$ have analytic solutions that are complicated functions of polylogarithms.  We adopt the efficient polylogarithm evaluation strategy following \cite{voigt2022comparisonmethodscalculationreal, voigt2023algorithmapproximaterealtrilogarithm} and \cite{ roughan2020polylogarithmfunctionjulia}, which enables rapid computation of polylogs of orders $2$, $3$, and $4$ with accuracies approaching machine precision.

\section{Tensors}
\label{sec:tensors}
\subsection{Algorithm}
\begin{figure}
    \centering
    \includegraphics[width=0.3\linewidth]{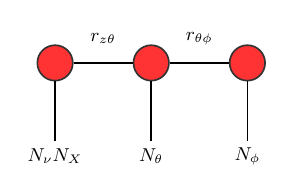}
    \caption{Graphical representation of the TT solution. \revtext{Each node corresponds to a tensor core whose dimensionality equals the number of connected edges. As shown in the figure, the specific intensity solution is decomposed into three tensor cores---$Z$ of order 2, $\Theta$ of order 3, and $\Phi$ of order 2---arranged from left to right.}}
    \label{fig:TT_sol}
\end{figure}

A key contribution of this paper is the development of a tensorized version of a ``traditional" MRT solver, which enables the solution of multigroup problems at discretizations that would be challenging (if not impossible) for conventional methods. The multigroup specific intensity solution has $N_X N_\nu N_\theta N_\phi$ unknowns, where $N_X$ is chosen represent the \textit{total} number of spatial coordinates ($= N_x$, $N_x N_y$, and $N_x N_y N_z$ for 1D, 2D, and 3D, respectively). $N_\nu$ represent the number of groups (i.e., frequency ``bins") and $N_\theta N_\phi$ is the total number of angles in a latitude/longitude ($\theta, \phi$) discretization, forming a 2D tensor product angular space.
We represent the specific intensity solution in a TT format as
\begin{align}
 \hat{I}_{k \ell p} = Z_{k} \Theta_{\ell} \Phi_{p}\label{eq:tensor_decomp_mer}.
\end{align}
Here, $Z \in \mathbb{R}^{N_X N_\nu \times \rtwo}$ represents a core that combines the spatial and frequency dimensions (i.e., a ``spatio-spectral" core), while $\Theta \in \mathbb{R}^{\rtwo \times N_\theta \times \rthree}$ and $\Phi \in \mathbb{R}^{\rthree N_\phi}$ represent the angular cores. $\rtwo$, $\rthree$ are the TT ranks. \revtext{Figure \ref{fig:TT_sol} presents a graphical depiction of the TT solution. The nodes, arranged from left to right, represent the TT-cores $Z$, $\Theta$, and $\Phi$. The number of edges connected to each node reflects the dimensionality of the corresponding core. Additional background on tensor networks can be found in \cite{biamonte2017tensor}.}

We refer the reader to \cite{gorodetsky25} for a review of the basic TT operations such as addition, integration, multiplication, and rounding.\footnote{\revtext{The solver implements 3 rounding methods-SVD \citep{oseledets2011tensor}, Gram \citep{al2022parallel}, and randomized \citep{al2023randomized}. The randomized rounding method increases each solution rank by some constant at each timestep during the randomized preconditioning step, followed by an SVD-based truncation using the same rounding tolerance as the standard SVD-based and Gram rounding methods.}} We re-emphasize that TT-operations assume compatibility between operands. For example, if wanting to add/multiply the solution $\hat{I}$ with a spatio-spectral quantity like the opacity $\kappa$, the operation requires both $\hat{I}$ and $\kappa$ to have the same format. Specifically, if $\hat{I}$ is represented as in Equation (\ref{eq:tensor_decomp_mer}), then the TT form of the opacity $\hat{\kappa}$ should be represented as \revtext{a rank-1 TT} $\Tilde{Z}_k \Tilde{\Theta}_l \Tilde{\Phi}_p$ where $\Tilde{Z} = \kappa$ and $\Tilde{\Theta}, \Tilde{\Phi}$ are filled with $1$ in order to artificially increase the dimensionality of the opacity. Any discrepancy in the formats of operands needs to be resolved via preprocessing before carrying out the operation.

All flux stencils studied in \cite{gorodetsky25} can be straightforwardly extended to the multigroup setting. The upwind flux and Rusanov fluxes \citep[as presented in][]{gorodetsky25} are independent of the local group opacity and therefore follow the same implementation as their gray counterpart.  The HLL interface flux is a function of the local optical depth $\tau_{c,f}$ that is a function of \textit{both} space and frequency, however, because we have opted for a \textit{merged} spatio-spectral core, tensor arithemetic in constructing the HLL flux is immediately straightforward (i.e., the tensor formats are compatible).  Alternative, more advanced tensor decompositions may not benefit from such simplicity (see \S \ref{sec:tests}).
The source solver presents similar considerations as the HLL flux operator. In evaluating the advanced specific intensity $I^{m+1}$, $\kappa_{a, f}$, $\kappa_{s, f}$, and $\varepsilon_f^\dagger$ equally vary with space and frequency. Consequently, addition and multiplication in the implicit update benefit from the unified spatio-spectral core employed in this work.

Nonetheless, we concede that our representation does not fully leverage separability between the spatial and spectral dimensions, which may manifest as additional low-rank structure within the core $Z$. This internal structure could be leveraged for greater compression for certain multigroup problems. We quantify the internal structure with the help of internal ranks and study their behavior in \S \ref{sec:discussion}. To summarize, the multigroup TTTT solver advances the TT solution via a step-then-truncate algorithm. \revtext{In this setting, the transport operator is advanced first, followed by a rounding step, then the source-term update, and finally another rounding operation to control rank growth.}

\subsection{Performance and cost}
We define the compression ratio
\begin{align}
    \mathcal{C} &:= \frac{N_\mathrm{trad}}{N_{TT}} = \frac{N_X N_\nu N_\theta N_\phi}{N_X N_\nu r_{z\theta} + N_\theta r_{z\theta}r_{\theta\phi} + N_\phi r_{\theta\phi}}\label{eq:compression}
\end{align}
which compares the traditional multigroup \revtext{solution size} to the TTTT one. Moreover, we define the speedup $\mathcal{S}$ as the ratio of the time taken by a traditional solver to the time taken by the TT-solver.\footnote{For the traditional solver baseline, we adopt the same performance metric from \cite{gorodetsky25}, but now run on a Mac \revtext{Studio with an M3 Ultra chip and $256$ GB of RAM} --- vacuum transport: $5.0 \times 10^8$, scattering: $3.7 \times 10^8$, absorption/emission: $3.6 \times 10^8$ degrees of freedom updates per second.  We assume perfect scaling of the gray solver with number of groups---an aggressive upper limit given multigroup introduces nontrivial Planckian integral evaluations.}

\begin{revblock}
The reported performance metrics compare the TTTT solver against a traditional solver using the same discretization. However, TTTT introduces additional timestep-wise errors that are controlled by a user-specified rounding tolerance. In total, the solver is subject to five primary sources of error: (1) spatial truncation error, (2) angular discretization error, (3) temporal discretization error, (4) frequency resolution error, and (5) tensor rounding error inherent to the TT representation. TT rounding can be viewed as behaving similarly to the arithmetic rounding inherent to floating-point numbers. We refer the reader to \cite{gorodetsky25} for a more detailed discussion of these error components. In principle, the rounding tolerance can be set close to machine precision to match the accuracy of traditional solvers, though this reduces achievable compression and computational speedup. As a consequence of rounding, conserved quantities are preserved up to the specified tolerance rather than to machine precision. Overall, TTTT provides a rigorous mathematical framework for controlling additional approximation error while exploiting structural properties of the solution.

In general, TT representations reduce storage requirements from $O(N^d)$ to $O(dNr^2)$ and computational costs from exponential dependence on $d$ to $O(dNr^3)$, with rounding typically being the dominant cost operation. This scaling assumes that each dimension associated with a TT core is discretized at a comparable resolution of $N$. In our solver, however, we typically have $N_x N_\nu \gg N_\theta, N_\phi$, so storage is dominated by the merged core $Z$ which has a size of $N_x N_\nu r_{z\theta}$. Similarly, TT-rounding is dominated by operations on $Z$, scaling approximately as $O(N_x N_\nu r_{z\theta}^2)$. This places practical limits on how finely space and frequency can be simultaneously discretized. Nevertheless, this design offers important advantages. Flattening the Cartesian spatial dimensions into a single core facilitates extension of TT solvers to an adaptive mesh refinement (AMR) framework. In addition to the implementation-related considerations discussed in the previous subsection, employing a TT core that merges the spatio-spectral dimensions offers advantages in problems with strong opacity-driven coupling between space and frequency. In particular, this representation helps mitigate pointwise artifacts in angle-integrated quantities such as $J$ and $E$, as discussed later.

\paragraph{Note on compression $\mathcal{C}$}
The TT solution used to compute the compression factor $\mathcal{C}$ in this work is evaluated after rounding. This does not capture the peak memory usage of the solver. In practice, memory usage is highest immediately after the transport update and before the subsequent rounding step, effectively increasing the post-rounding storage by a factor $k_s$. The exact value of $k_s$ depends on several factors, including the rounding algorithm, rounding frequency, discretization, and tensor ranks.  In the context of this study—assuming that the merged core $Z$ dominates storage and that TT ranks remain relatively stable across timesteps—we estimate a worst-case range of $k_s = 3-6$ for 1D problems and $5-12$ for 2D problems. A more detailed and general discussion is provided in Appendix \ref{sec:appendix}. These intermediate storage estimates are implementation-dependent and if memory usage becomes prohibitive, it can be reduced by forming partial sums and performing more frequent rounding within the transport step, potentially lowering $k_s$ to values as small as $2$. We do not make definitive claims regarding the impact of $k_s$ on the reported compression values and instead focus on its effect on TT storage. This is because compression is defined relative to the baseline storage $N_X N_\nu N_\theta N_\phi$, rather than accounting for implementation-specific overheads (i.e., a comfortable implementation of RK1 might even invoke two registers). While a more rigorous comparison of memory usage between TT and traditional solvers is warranted, we do not pursue it in this work. Instead, we rely on the reported compression values as reasonable estimates of the practical memory savings.

\end{revblock}

\section{Test Problems} \label{sec:tests}
 Building on the results of \cite{gorodetsky25}, we anticipate that the TTTT solver will achieve substantial compressions and speedups relative to ``traditional" multigroup $S_N$ implementations. Herein, for representative multigroup test problems, we assess the rank structure of the tensor decomposition employed in this work. We introduce some metrics associated with the internal structure of $Z$ in \S \ref{sec:discussion} which along with $\mathcal{C}$ and $\mathcal{S}$ dictate the performance of our solver and are reported for all test problems alongside the rounding \revtext{algorithms\footnote{\revtext{The SVD-based algorithm is used for most problems to ensure the accuracy of our rank study while Gram rounding is employed for the stellar irradiation problem to accelerate the solver.}} and} tolerances inside Table \ref{tab:conclusion}.

We adopt custom unit systems for all tests except Graziani’s test (which enrolls cgs units). In the multigroup tests, these modified units are taken from the corresponding gray versions described in \cite{gorodetsky25}. We note that the choice of unit system can affect solution accuracy, potentially due to floating-point effects, and as with all numerical methods, a unit system where the magnitude of the solution is of order unity is ideal.

 \revtext{Lastly, following \cite{gorodetsky25}, we adopt a uniform spatial mesh and a latitude/longitude angular discretization with $N_\phi = 2 N_\theta$ with quadrature points uniformly sampled in $\cos(\theta)$ and $\phi$. Unless otherwise specified, we consider $\nu_1, \nu_2, \dots \nu_{N_\nu - 1}$ to be logarithmically spaced, with $\nu_0 = 0$ and $\nu_{N_\nu} = \infty$.}

\subsection{Multigroup Hohlraum}
We begin by assessing the multigroup TTTT solver's performance in the free streaming limit via the 2D hohlraum vacuum transport problem \citep{mocmc, white23, gorodetsky25}.  A square spatial domain \revtext{$[0, 2] \times [0, 2]$} is discretized into $N_X = 128^2$ cells.  We choose $N_\theta \times N_\phi = 512\times1024$ angles to reduce errors from ray effects and invoke an upwinded flux. Fixed left and bottom isotropic boundary sources with frequency integrated specific intensity $I=J=1$ send radiation propagating throughout the domain.

We adapt the problem to multigroup as follows.  The frequency domain is discretized into $N_\nu = 10$ bins. The isotropic boundaries at the left and bottom are initialized in two ways (1) $I_f = J_f = 1/N_\nu$ and (2) $I_f = J_f = \varepsilon_f (T_\mathrm{init})$ such that $a_r T_\mathrm{init}^4 = 4\pi/c$.  The former setup sets each group identically, and each group should evolve identically; the second setup will have each group propagate identically, but with a magnitude set by the initial thermal spectrum at the active boundaries.  Both setups have an analytic result derived from the gray solutions in \cite{white23} and \cite{gorodetsky25}, which we do not repeat here. \revtext{ We note that the numerical solution is compared with analytics over $x, y \in [0, 1]$ to account for the assumption of infinite isotropic boundaries inherent to the analytic solution. }

\begin{figure}
    \centering
    \includegraphics[width=0.6\linewidth]{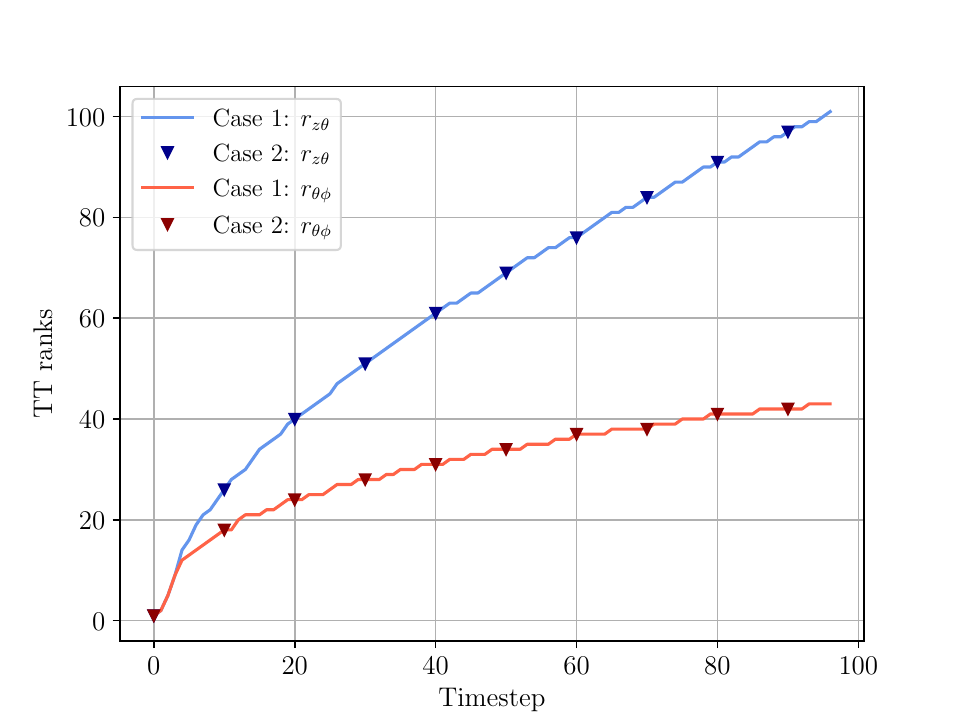}
    \caption{Rank evolution of the multigroup hohlraum problem for both cases. The plot reveals that the rank evolution stays the same for both cases despite the discrepancy in initialization. We attribute this to the fact that the solution in every bin evolves exactly according to the gray solution scaled by the initial intensity in that bin. This leaves the TT-ranks unchanged for arbitrary intensity initializations across the frequency domain.}
    \label{fig:hohlraum_ranks}
\end{figure}

We explore both configurations to demonstrate interesting properties of our solver. We plot the TT-rank evolution in Figure \ref{fig:hohlraum_ranks} and observe that both initializations present identical rank evolution even though the distributions of specific intensity across the frequency domain are different.  This stems from the fact that frequency only influences the solution through a group-dependent constant that scales the gray solution. Consequently, the rank evolution for the chosen discretization should remain unchanged even if the intensities in different groups are initialized randomly.\footnote{This assumes spatially uniform and groupwise-identical left and bottom boundary conditions. For instance, initializing the left and bottom boundaries with distinct values of $T_{\mathrm{init}}$ would lead to a different TT-rank behavior compared to our two included cases.}

\begin{figure}
    \centering
    \includegraphics[width=1.0\linewidth]{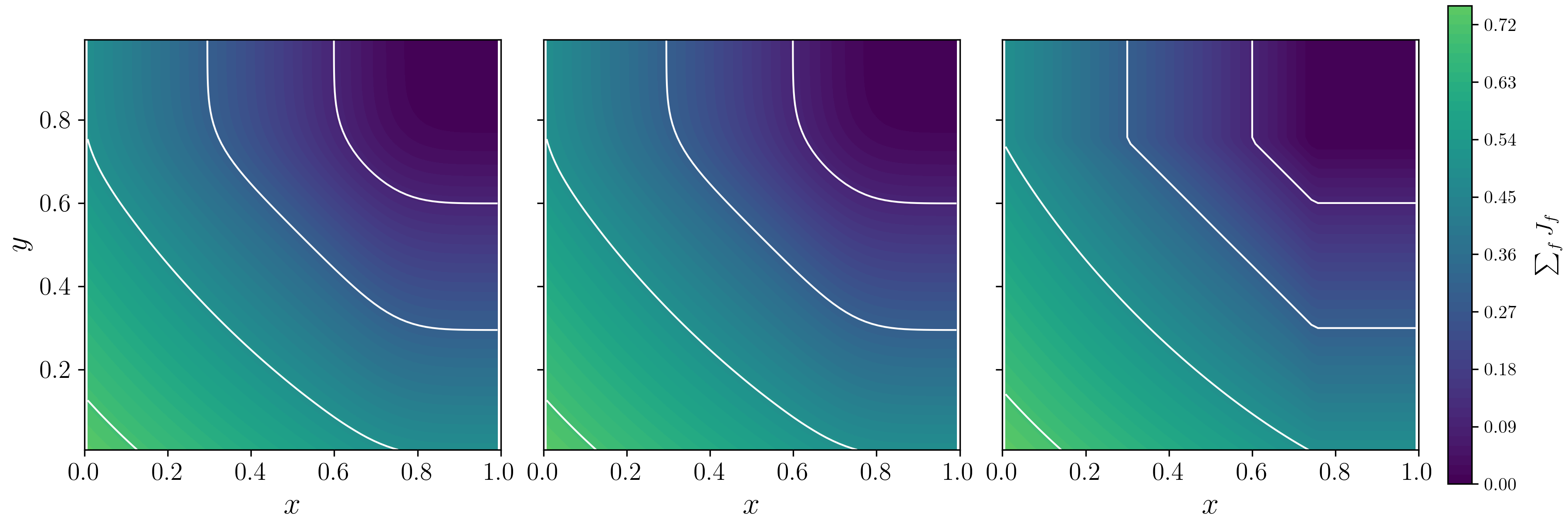}
    \caption{Reduction of multigroup mean intensities to the gray case for case 1 (left) and 2 (center) plotted alongside the corresponding analytics (right) at $t=0.75$. Plots are overlayed with white contour lines at $\sum J_f = [0.1, 0.3, 0.5, 0.7]$. The numerical solution follows the analytical solution closely with compression and speedup amounting to $\sim$\revtext{4565} and $\sim$\revtext{60} respectively.}
    \label{fig:hohlraum}
\end{figure}

Figure \ref{fig:hohlraum} compares the analytical and numerical solutions for $\sum_fJ_f$ for both cases at $t=0.75$ \revtext{(CFL = 0.5)}. For both cases, the TTTT solver recovers the gray analytics (albeit exhibiting large 1st order spatial truncation errors) and presents a total compression ratio of \revtext{$\mathcal{C} \approx 4565$} and a speedup of $\mathcal{S} \approx 60$.

Finally, we remark that the decoupling of frequency dimension in the solution hints at rank-1 structure inside $Z$ and suggests potential for additional compression gains by separating the frequency core (see later discussion in \S \ref{subsubsec:core_structure}).

\subsection{Thermal Relaxation}
Moving beyond vacuum transport, the multigroup thermal relaxation test problem investigates a frequency-dependent problem invoking the absorption and emission source terms and radiation feedback \citep{jiang22}. In this test, a material and radiation are initialized to different temperatures ($T$ and $T_r$) inducing energy exchange among them until thermal equilibrium is reached.  Each group evolves on a timescale determined by its group opacity $\kappa_{a,f}$.  The final equilibrium temperature $T_\mathrm{eq}$ is independent of group structure $\nu_f$ and opacity structure $\kappa_{a,f}$, and is derived from total energy conservation \citep[see, e.g., ][]{gorodetsky25}.

We consider $\rho=1$, $c_v = 8$, $T_\mathrm{initial} = 10$, $T_{r,\mathrm{initial}} = 1$ and adopt a unit system wherein $a = c= 1$.  Note that the resulting equilibrium temperature for these settings is $T_\mathrm{eq} \simeq 2.77$.  We choose $N_X = 4$ cells to span the spatial domain of extent $[0, 4]$. We use $N_\theta \times N_\phi = 512 \times 1024$ angles and initialize the mean intensity for each group as $J_f = \varepsilon_f(T_{r,\mathrm{initial}})$. We highlight that the choice of spatial and angular discretization is immaterial, since the solution in each group remains isotropic and spatially uniform, reducing the problem to one that depends solely on frequency. Although the numerical flux choice is inconsequential, we invoke the Rusanov flux with $S^+= c$ \revtext{and set CFL = 0.1} for this test.

\subsubsection[\texorpdfstring{$N_\nu = 3$}{N\_nu = 3}]{$N_\nu = 3$}
We test the solver for $N_\nu=3$ with $\nu_1 = k$ and $\nu_2 = 2k$ where $k=1.5 \times 10^{11}$. We set $\kappa_{a,1} = 10, \; \kappa_{a,2} = 1, \; \kappa_{a,3} = 0.1 $ and run the solver to $t_\mathrm{final}=40$. We normalize the radiation energy density by $aT_{eq}^4$ and denote it by  $\overline{E}_f$.

\begin{figure}
    \centering
    \includegraphics[width=0.5\linewidth]{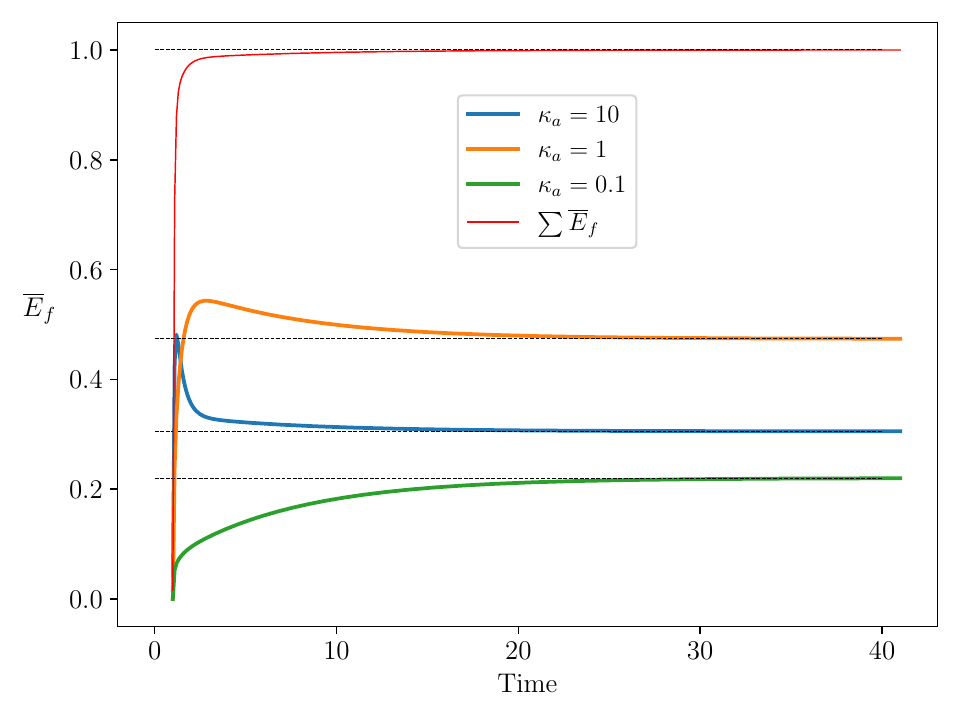}
    \caption{Plot of the normalized radiaiton energy $\overline{E}_f$ v/s time. Different colored lines represent the solution evolution in different frequency groups and each plot converges to analytics as thermal equilibrium (at $T = T_{eq}$) is achieved.}
    \label{fig:thermal_relax3}
\end{figure}

The analytic equilibrium states for this problem are \revtext{$\overline{E}_1 = 0.31$, $\overline{E}_2 = 0.47$, and $\overline{E}_3 = 0.22$.} The numerical solution is presented in Figure \ref{fig:thermal_relax3} where we plot the time evolution of $\overline{E}_f$ for $f=1, 2, 3$ along with $\sum_f\overline{E}_f$. The solutions converge to and match the analytics in every group signifying that the solver works as expected. We observe that the path to equilibrium varies across groups. Groups with higher opacities exhibit rapid changes in $\overline{E}_f$ during the early stages and dominate the temperature evolution of the medium at these times. As the temperature evolves, additional groups become active.

\subsubsection[\texorpdfstring{$N_\nu = 100$}{N\_nu = 100}]{$N_\nu = 100$}
\begin{figure}
    \centering
    \gridline{
        \fig{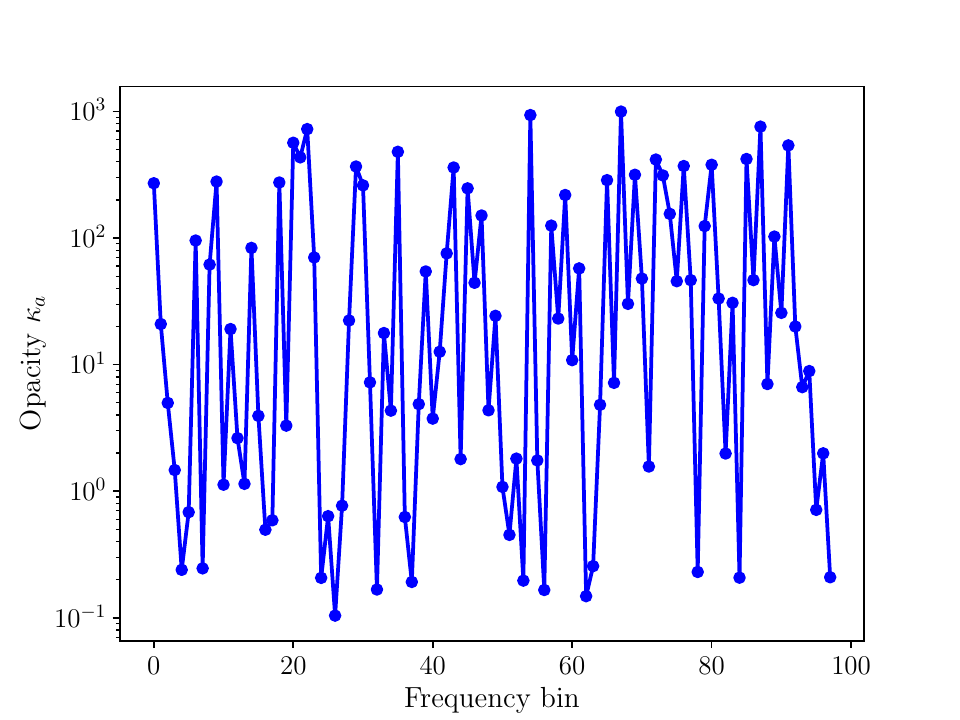}{0.49\textwidth}{(a)}
        \fig{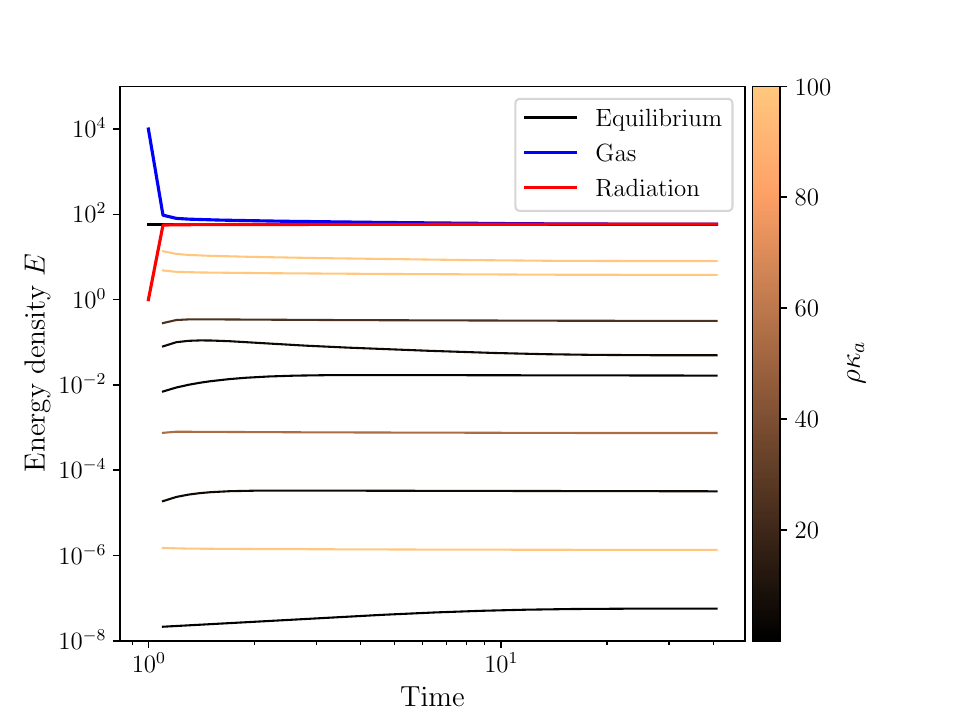}{0.49\textwidth}{(b)}
    }

    \caption{
        (a) Opacities $\kappa_{a,f}$ for the $N_f = 100$ relaxation test. 
        (b) The variation of thermal energy density of the material ($\rho c_v T$) (blue), radiation energy density for some of the groups (colorbar), 
        and the total radiation energy density (red) with time. 
        The numerical solutions converge to the predicted analytic equilibrium. 
        Notice that the radiation energy density generally reaches a steady value slower for groups with lower opacities 
        (the darker lines flatten later compared to the orange ones).
    }
    \label{fig:therm100}
\end{figure}

To test the robustness of the implementation, we consider a case with $N_\nu = 100$ \revtext{and $(\nu_1, \nu_{99}) = (10^{6}, 10^{15})$}. The group opacities are assigned using a random number generator such that $\kappa_{a,f} = 10^x$, where $x$ is uniformly sampled from the interval \revtext{$[-1, 3]$} (see Figure \ref{fig:therm100}a). The system is evolved up to time $t_\mathrm{final} = 10$.

\begin{figure}
    \centering
    \includegraphics[width=0.6\linewidth]{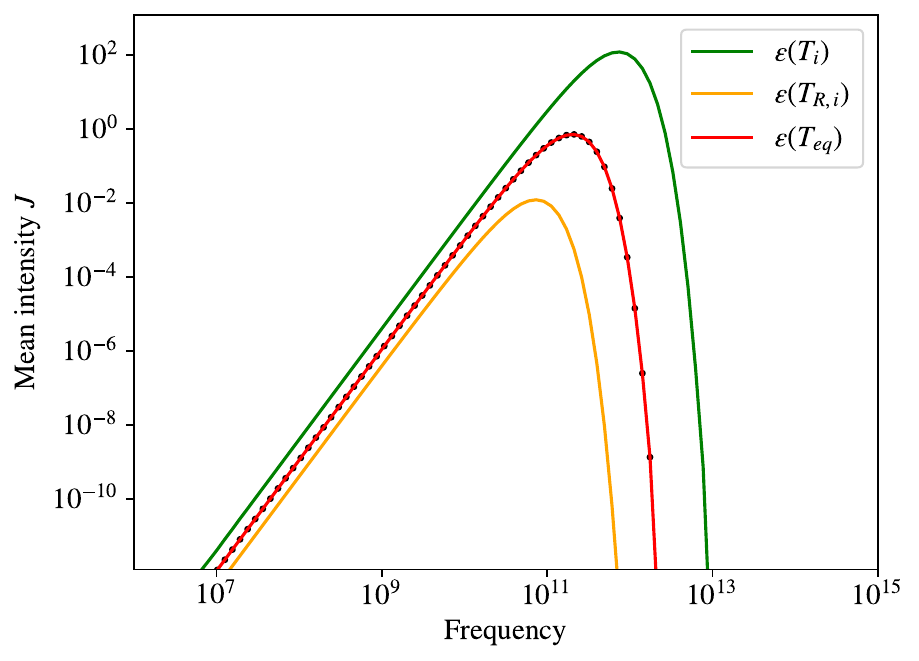}
    \caption{\revtext{Solution at the final timestep for the thermal relaxation problem with $100$ frequency groups, shown as a $J$ versus frequency spectrum. The scatter points denote the mean intensity in each group, which initially follow a Planck distribution evaluated at the initial radiation temperature. Over time, the groups relax toward the Planck distribution corresponding to the equilibrium temperature at a rate determined by the group opacity $\kappa_{a,f}$. A $12$-second animation illustrating the evolution of these groups is available in the HTML version.}}
    \label{fig:eps_video}
\end{figure}

We visualize the solution in the included movie as a time-dependent spectrum, plotting $J_f$ versus frequency $\nu$. Initially, the material and radiation spectra are given by $\varepsilon_f(T_\mathrm{initial})$ and $\varepsilon_f(T_{r,\mathrm{initial}})$, respectively. As the system evolves, both spectra relax toward a Planck distribution corresponding to the equilibrium temperature $T_\mathrm{eq}$, i.e., $\varepsilon_f(T_\mathrm{eq})$ \revtext{(see Figure \ref{fig:eps_video})}. We observe each group evolving at a rate proportional to its respective opacity $\kappa_{a,f}$ (albeit each group must settle before $T_\mathrm{eq}$ is achieved).

Alternatively, the solution can also be visualized as an energy versus time plot as shown in Figure \ref{fig:therm100}b. We plot the evolution of gas energy (blue) and total radiation energy (green). The radiation energy in various bins is plotted and colored according to the value of opacity in the bin. The dynamics here are fundamentally the same as in the previous test, though they now appear far more complex. Even still, the solver captures the temperature variation for each group and demonstrates perfect compression. At all times, the thermal relaxation solution admits a rank-1 TT representation, yielding compressions of $\mathcal{C} = 4000$ when $N_\nu=3$ and $\mathcal{C} = 1.2 \times 10^5$ when $N_\nu = 100$. 

The extent to which compression improves with increased spectral resolution depends on both the TT ranks and the sizes of the TT cores. The thermal relaxation problem represents a best-case scenario with respect to the former, as it yields a rank-1 solution. Consequently, the substantial increase in compression observed when increasing from $N_\nu = 3$ to $N_\nu = 100$ is not generally applicable to other problems or even to different resolutions of the same problem. In cases that produce higher-rank solutions, the gain in $\mathcal{C}$ with increasing $N_\nu$ is more modest. Moreover, for the present problem, this increase would be significantly smaller if $N_X N_\nu \gg N_\theta, N_\phi$.

\subsection{Gaussian diffusion}
In this problem, an initial 1D Gaussian pulse is diffused at a rate inversely proportional to the scattering opacity of the medium. To extend this to multigroup, we simply assign different scattering opacities to different groups and probe the TT-ranks of the solution over time. The analytics for the radiation energy density are given by
\begin{equation}
    E_f(t,x; t_0) = \frac{A}{ \sqrt{4 \pi D_f (t_{0,f} + t)}} \exp \left[- \frac{x^2}{4 D_f (t_{0,f} + t)} \right],\label{eq:diffusion}
\end{equation}
where $A$ denotes the initial Gaussian amplitude, and both the initial time $t_{0,f}$ and the scattering diffusion coefficient $D_f := c / (3 \rho \kappa_{s,f})$ can be group dependent \citep{sekora10, jiang14, jiang21}.

\begin{figure}
    \centering
 \includegraphics[width=1.0\linewidth]{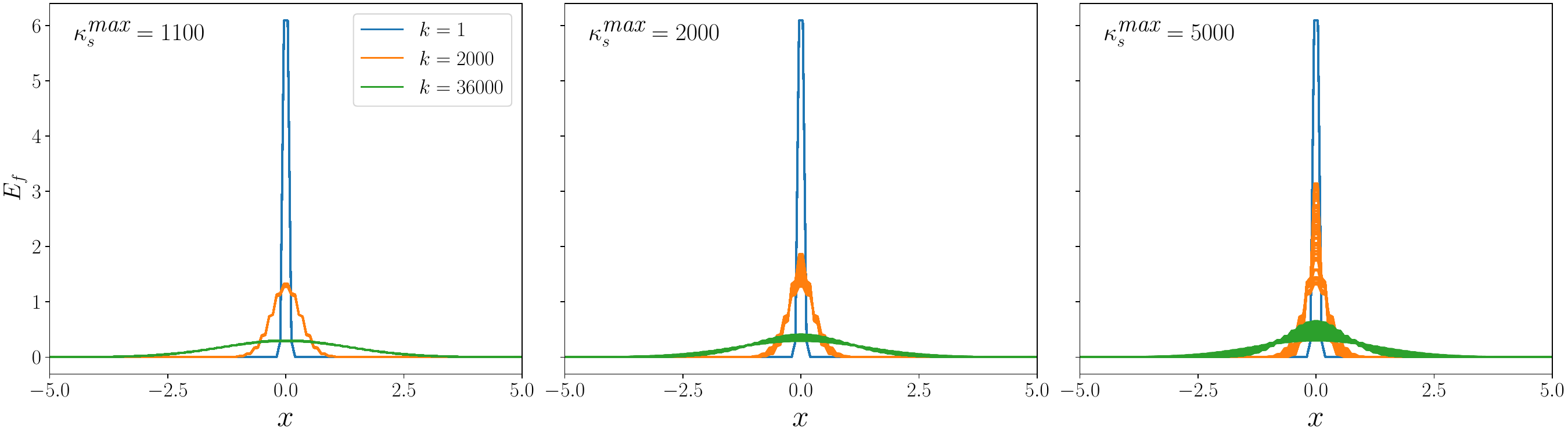}
    \caption{Radiation energy density snapshots for \textit{all} frequency groups at timesteps $k = 1, 2000,$ and $36000$ (blue, orange, and green, respectively), shown for increasing values of $\kappa_s^{max}$ from left to right. Initially, the solution is identical across groups, producing an overlapping set of Gaussian profiles. As time evolves, the groupwise solutions diverge due to differences in $\kappa_{s,f}$, leading to progressively dissimilar profiles at a given timestep, until they asymptotically decay toward zero. The degree of dissimilarity among the solutions is reflected in the ``thickness" of the profiles, which is directly determined by the range of $\kappa_{s,f}$ present in the problem.}
    \label{fig:diffusion}
\end{figure}

We initialize an isotropic field $I_f(t=0, x, \nu, \theta, \phi) = J_f(t=0, x; t_0 = t_{0,f})$ and allow it to diffuse. We set \revtext{$A=1$ and $t_{0,f} = 10^{-3}/D_f$} to ensure identical initial conditions for every group. We set constants $\rho$, $A$, and $c$ to $1$ each and use $128$ spatial cells distributed in $[-5, 5]$, $100$ frequency bins, and $1024 \times 2048$ angles to simulate the problem \revtext{(CFL = 1)}. Since this problem is positioned well into the diffusive regime, we evolve the transport operator via the Rusanov flux with a wavespeed $S^+=0$ (i.e., a pure finite difference). For the test, the scattering opacities $\kappa_{s,f}$ are sampled uniformly from $[1000, \kappa_s^{\textit{max}}]$. We study the impact of expanding the range of the uniform distribution on the TT ranks by varying $\kappa_s^\textit{max}$. Increasing this range leads to greater spatial variability in the solution across different frequency groups. \revtext{This behavior can be observed in Figure \ref{fig:diffusion}.}

The multigroup Gaussian diffusion problem serves not only as a validation test for the correct implementation of the scattering term in the solver, but also as an important benchmark for assessing the applicability of the multigroup formulation. In the previous tests, the radiation energy density could be written as a simple product of a frequency-dependent factor and a spatially dependent factor. In contrast, in Equation \eqref{eq:diffusion} the spatial and frequency dimensions are more strongly coupled, suggesting an internal rank of $Z$ greater than one. This coupling makes the problem interesting both in terms of the internal structure of $Z$ and its implications for the resulting TT ranks.

\begin{figure}
    \centering
    \includegraphics[width=0.5\linewidth]{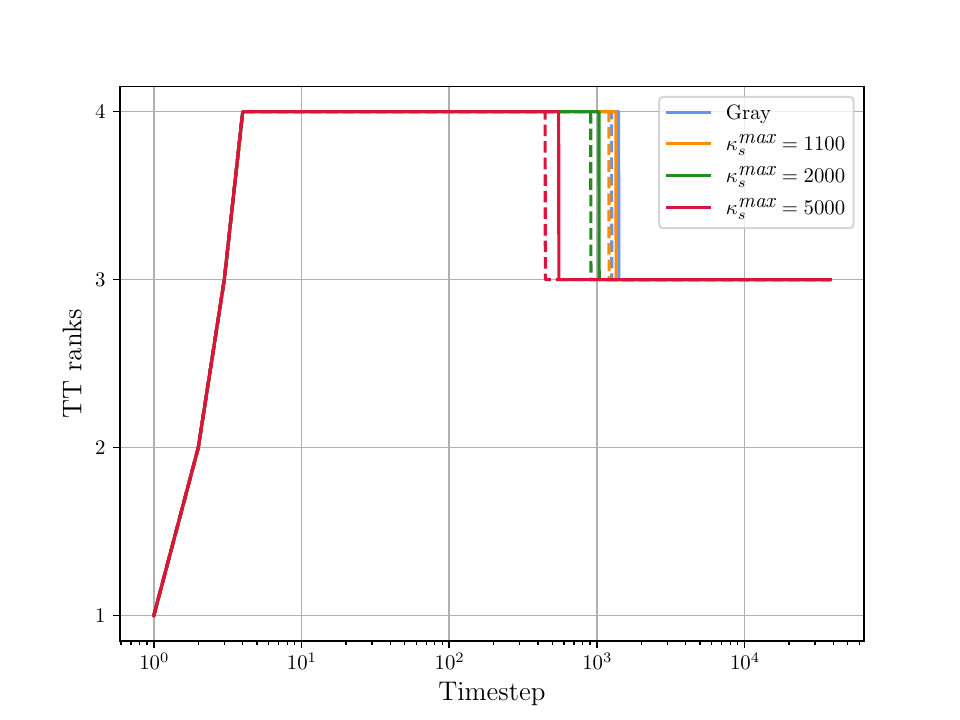}
    \caption{Evolution of TT-ranks for $\kappa_s^{\textit{max}}=[1000, 1100, 2000, 5000]$. The rank $\rtwo$ is shown with solid lines and $\rthree$ with dashed lines. The ranks initially increase and then decrease as the solution approaches isotropy. Both the rank evolution and compression behavior remain similar across all cases. The minimum compression occurs when the TT ranks are largest and is $\sim$$3.5 \times 10^5$.}
    \label{fig:merged_diffusion}
\end{figure}

We find that the TT ranks are largely insensitive to changes in spatio-spectral complexity. This behavior is illustrated in Figure \ref{fig:merged_diffusion}, which shows the evolution of the TT ranks for $\kappa_s^\textit{max}=1000,1100,2000$, and $5000$.\footnote{Note that $\kappa_s^\textit{max}=1000$ reduces the multigroup problem to a gray problem with uniform scattering opacity.} For each value of $\kappa_s^\textit{max}$, TT ranks $\rtwo$ and $\rthree$ are plotted using the same color, with $\rthree$ decreasing first. Although the TT ranks drop at different times, their overall behavior remains similar. We note that this behavior is expected provided that spatio-spectral complexity does not strongly influence the angular dynamics (i.e., the scattering source term maintains $\sim$isotropy throughout each group's evolution). Since our TT representation does not exploit separability between the frequency and spatial dimensions, it does not invite rank growth due to the strong space-frequency coupling in this problem---that is, we observe low rank structure as the solution is $\sim$isotropic over the full spatio-spectral space.  The minimum compression observed for this problem is $\mathcal{C} \approx 3.5 \times 10^5$ and the speedup achieved is $\mathcal{S} \approx 5865$.

\subsection{Multigroup Graziani prompt spectrum (slab version)}
The multigroup Graziani problem serves as an excellent benchmark for multigroup implementations and is typically run in 1D spherical geometry \citep[see, e.g.,][]{graziani2008computational, jiang22}. In this work, we opt for its planar counterpart (i.e., the ``slab" version) which equally has analytic solutions \citep{graziani2008computational}.  This subsection is divided into two main parts. We initially gauge the solver for accuracy and performance on the prompt spectrum problem as described in \cite{graziani2008computational}. In the second part, we test the limits of the solver by adding more complexity to the original problem.

\subsubsection{Original formulation}\label{sec:Graziani_trad}

Graziani's prompt spectrum problem considers an opaque slab with a predefined opacity structure (that of brominated plastic). A boundary condition set to a radiation temperature of $T_s=0.3$ keV acts on a background with an initial material and radiation temperature of $T_c = 0.03$ keV. 50 frequency groups are used to divide the frequency domain with $\nu_1 = 0.003$ keV and $\nu_{51} = 30$ keV. The medium density is prescribed as $0.0916$ $\text{g}\mkern3mu \text{cm}^{-3}$ and the specific heat capacity is set to a exceedingly large value to effectively maintain a constant medium temperature throughout the course of the simulation. We monitor the distribution of radiation energy density among all the different frequency groups at a specific point in space which we call the ``fiducial point." In the examples simulated, we use $N_X = 120$ cells and $N_\theta \times N_\phi = 64 \times 128$ angles and consider a point located at a distance of \revtext{$2.115 \times 10^{-2}$} cm from the left boundary to be our fiducial point. All tests in this subsection invoke the HLL flux with $\beta = 20$ \revtext{and CFL = 0.1} \citep[see][]{gorodetsky25}.
 
\begin{figure}
    \centering
    \includegraphics[width=1.\linewidth]{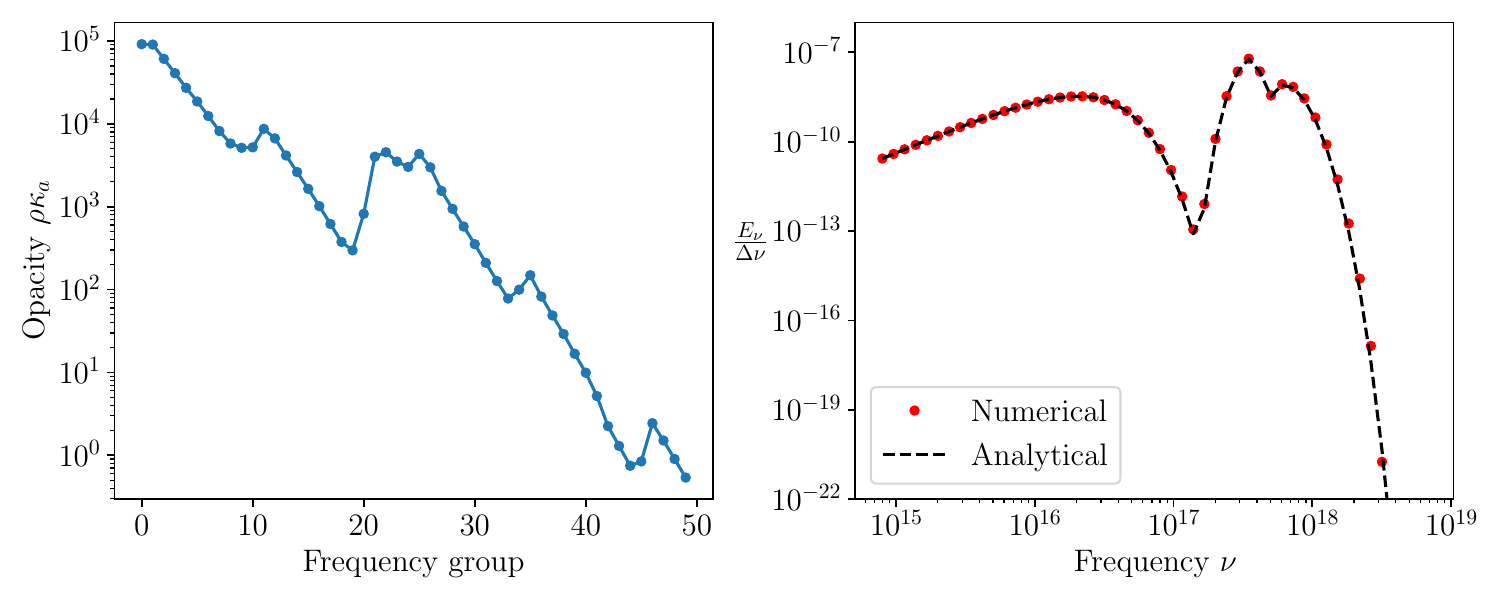}
    \caption{Brominated plastic opacity profile used in the \cite{graziani2008computational} prompt spectrum problem (left) and comparison between the analytical and TTTT solutions (right). The two peaks observed in the solutions correspond to the maxima of the Planckian emissions from the medium and the boundary. The TTTT solver accurately reproduces the analytic result while achieving compression gains of approximately $296\times$.
}
    \label{fig:graziani_result}
\end{figure}

The initial energy distribution at the fiducial point is isotropic and takes the shape of a Planckian evaluated at temperature $T_c$. Since the medium is in thermal equilibrium with the radiation field, the energy distribution remains unchanged until the information from the intensity source at the left boundary reaches it. The energy distribution will eventually converge to the profile as displayed in Figure \ref{fig:graziani_result}. The first peak arises from the Planckian distribution at the medium temperature $T_c$, representing thermal emission from the medium itself. The second peak corresponds to the Planckian peak of the source at temperature $T_s$. Energy at the frequency associated with the source's peak is able to reach the fiducial point more effectively due to the lower opacities at frequencies higher than that peak.
  
The numerical TT solution was obtained using a rounding tolerance of $10^{-6}$ and we observe that it captures the analytics accurately while giving compressions and speedups of $\mathcal{C} \approx 296$ and $\mathcal{S} \approx 1.5$.

\subsubsection{Spatially varying opacities}
Mediums in practical problems might have a mixture of materials scattered throughout space adding more complexities for our solver to resolve. In this test, we set up the Graziani problem with different artificial materials distributed in space. Each material is characterized by a unique opacity structure and as a result, the opacity field varies not only with frequency but also spatially. The frequency dependence follows a power-law trend common to all materials, with random oscillatory components introduced to distinguish individual materials---that is, we logarithmically decrease $\rho \kappa_{a,f}$ from $[10^8, 0.5]$ over frequencies $[\nu_1, \nu_{N_\nu-1}]$ and then add group-wise oscillations with an amplitude sampled from a Gaussian distribution with zero mean and variance $(m \cdot \rho \kappa_{a,f})^2 / N_\nu$ where $m=5$. We vary the number of materials in the medium to consider different levels of complexity in the corresponding opacity fields. The test is conducted with 1, 10, 20, 40, and 120 materials distributed uniformly in space, and we examine the solver’s performance for different rounding tolerances. The discretization for these tests was set as $(N_X, N_\nu, N_\theta, N_\phi) \equiv (120, 102, 64, 128)$.

\begin{figure}
    \centering
    \includegraphics[width=1.0\linewidth]{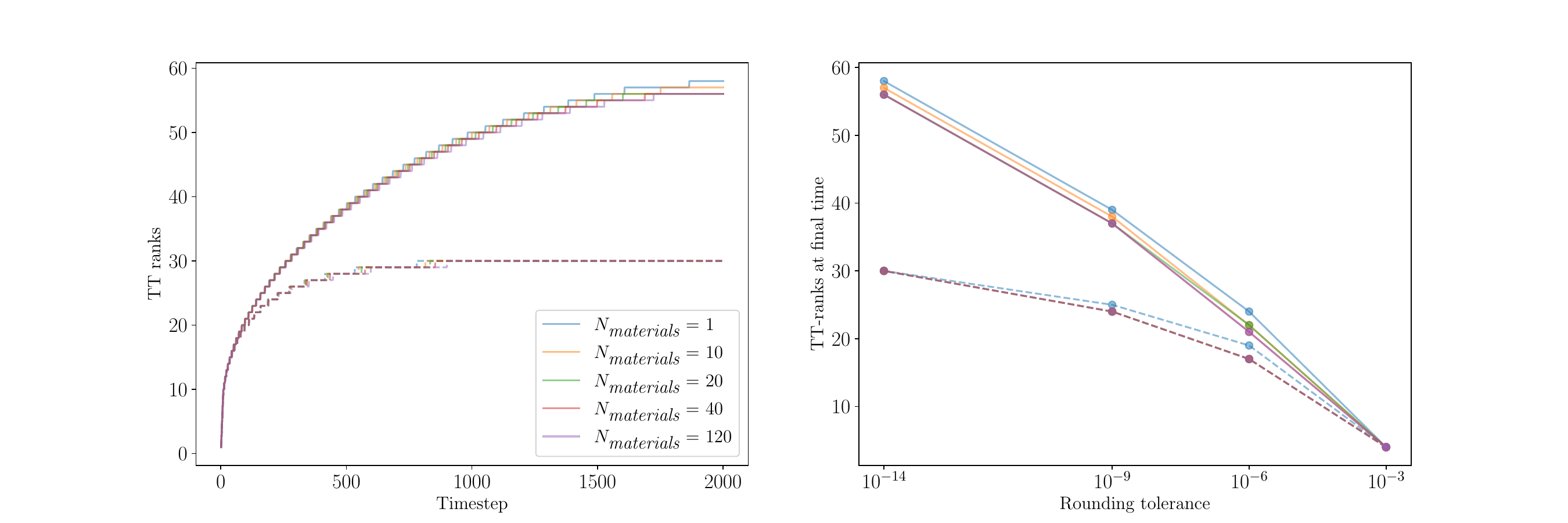}
    \caption{TT-rank evolutions for different $N_{materials}$ at a rounding tolerance of $10^{-14}$ (left), and final TT ranks as functions of the rounding tolerance for varying material counts (right). $\rtwo$ is shown with solid lines and $\rthree$ with dashed lines. The near-complete overlap of curves across different material counts indicates that, for the chosen tensor decomposition, the TT ranks are largely insensitive to spatio-spectral complexity arising from spatially varying opacities. The solver achieves significant compression relative to a conventional representation, with compression ratios $\mathcal{C}$ ranging from approximately $121$ at a rounding tolerance of $10^{-14}$, increasing to about $1985$ at a tolerance of $10^{-3}$.
}
    \label{fig:graziani_merged_ranks}
\end{figure}

Figure \ref{fig:graziani_merged_ranks} shows the evolution of the TT ranks for varying numbers of materials ($N_{materials}$) at a rounding tolerance of $10^{-14}$
  (left), as well as the final TT ranks as a function of rounding tolerance for different material counts (right). Rank $\rtwo$	
  is plotted with a solid line, while $\rthree$
  is shown with a dashed line. The curves for different material counts exhibit similar evolution, again indicating that the TT ranks for our selected tensor decomposition are largely insensitive to the spatio-spectral complexity introduced by spatially varying frequency-dependent opacities. Overall, the solver delivers substantial compression compared to a conventional solver representation, $\mathcal{C}$—with ratios of roughly $121$ at a rounding tolerance of $10^{-14}$, and rising to about $1985$ at a tolerance of $10^{-3}$. We also observe speedups ranging from $\mathcal{S} \approx 2$ for a rounding tolerance of $10^{-3}$ to $\mathcal{S} \approx 1.4$ for a tolerance of $10^{-14}$.

\subsection{Multigroup stellar irradiation}

We extend the the gray version of the stellar irradiation problem \citep{klassen14, gorodetsky25} to a multigroup setting. The spatial domain is $[-1000, 1000] \times [-1000, 1000]$ AU. We initialize a circular cold clump with a radius of $267$ AU at the origin and 2 radiating stars with radii $8$ AU each centered at [0, -500 AU] and [-500 AU, 0]. The clump and medium temperatures are set to $T=T_r =1$ K while the stars radiate at $T=T_r =140$ K. Additionally, the medium, clump, and star densities are assigned values of $10^{-20}$, $10^{-16}$, and $10^{-12}$ $\text{g} \mkern3mu \text{cm}^{-3}$ respectively.

The problem is run with 
$310^2$ spatial cells, 
$512 \times 1024$ angles, and 20 groups—exceeding a trillion total parameters. We initialize the group opacities so that $\kappa_a$ increases logarithmically from $10^{-6}$ at $f=1$ to $1$ at $f = N_\nu/2$, and then decreases logarithmically back to $10^{-6}$ by $f = N_\nu$. The frequency bounds are set to $(\nu_1, \nu_9) \equiv (10^{12}, 10^{14}) \; \mathrm{s^{-1}}$ so that the groups concentrate around the Planckian peak corresponding to the radiation temperature of the stars. Finally, the transport term is computed using the Rusanov flux with a wave speed of $S^+=c$ and the \revtext{CFL is set to 0.4}.

\begin{figure}
    \centering
    \includegraphics[width=1.\linewidth]{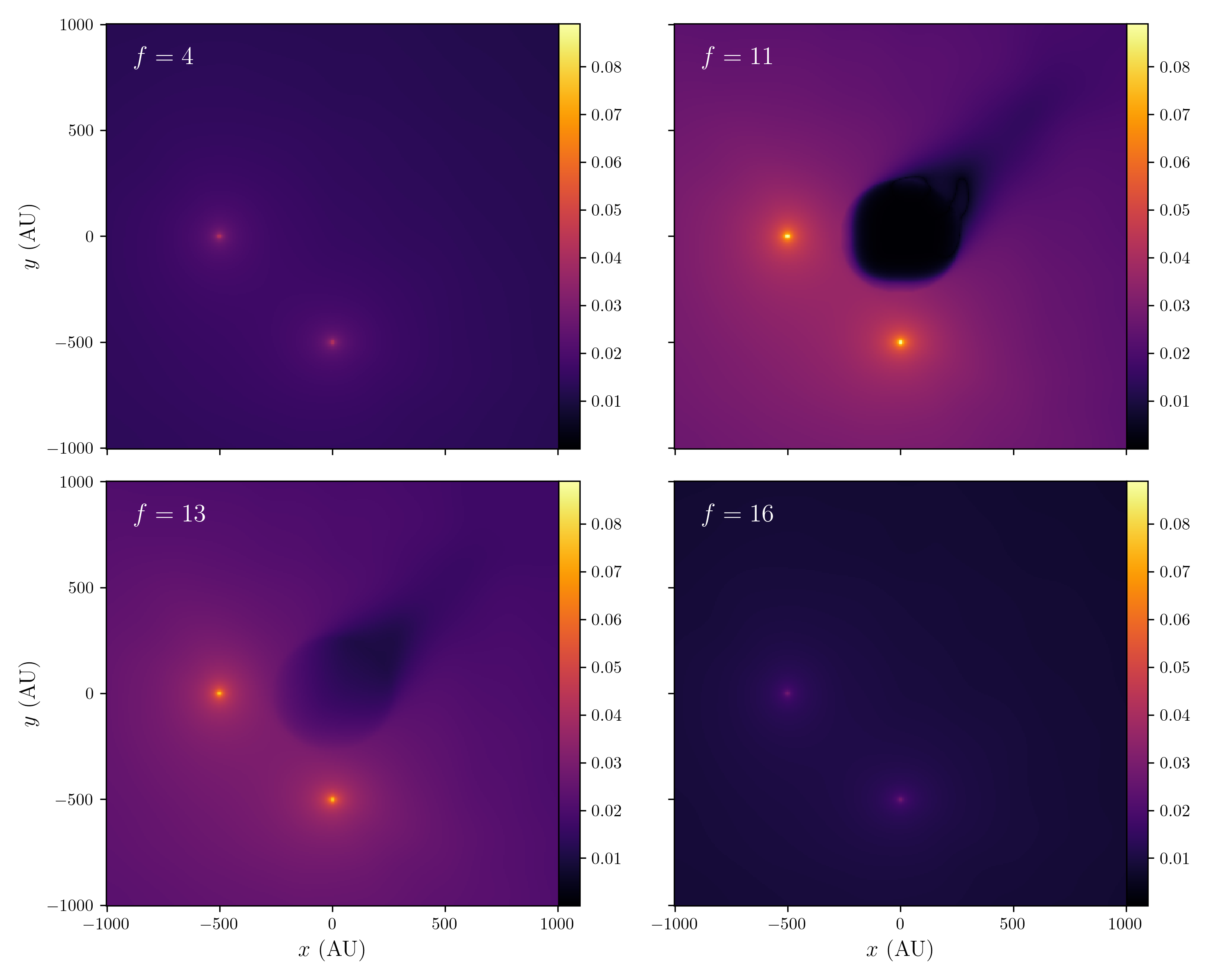}
    \caption{(From top-left in clockwise direction) $\propto E_f^{1/4}$ fields for frequency groups $f = 4$, $11$, $16$, and $13$. The clump opacity is the greatest for $f=11$ resulting in a more prominent shadow. Radiation is allowed to permeate as one moves further from the central bins. This can be observed in the form of partial permeation in case of $f=13$ and almost complete permeation for $f=4, 15$. We observe compressions and speedups of $\sim 4500 \times$ and $280 \times$ respectively in this problem.}
    \label{fig:multigroup_stellar}
\end{figure}

Figure \ref{fig:multigroup_stellar} plots $\propto E_f^{1/4}$ for groups $4$, $11$, $13$, and $16$ after a $\sim$steady state is achieved. For $f = 11$, the clump exhibits maximal opacity, leading to the most significant shadowing effect. Radiation penetration increases with distance from the central bins, appearing as partial permeation for 
$f=13$ and nearly full permeation for 
$f=4$ and $f=16$. The problem yields $\mathcal{C}$ and $\mathcal{S}$ on the order of $4500 \times$ and $232 \times$ respectively.

\section{Discussion} \label{sec:discussion}

\subsection{Binwise compression}
\begin{figure}
    \centering
    \includegraphics[width=0.5
\linewidth]{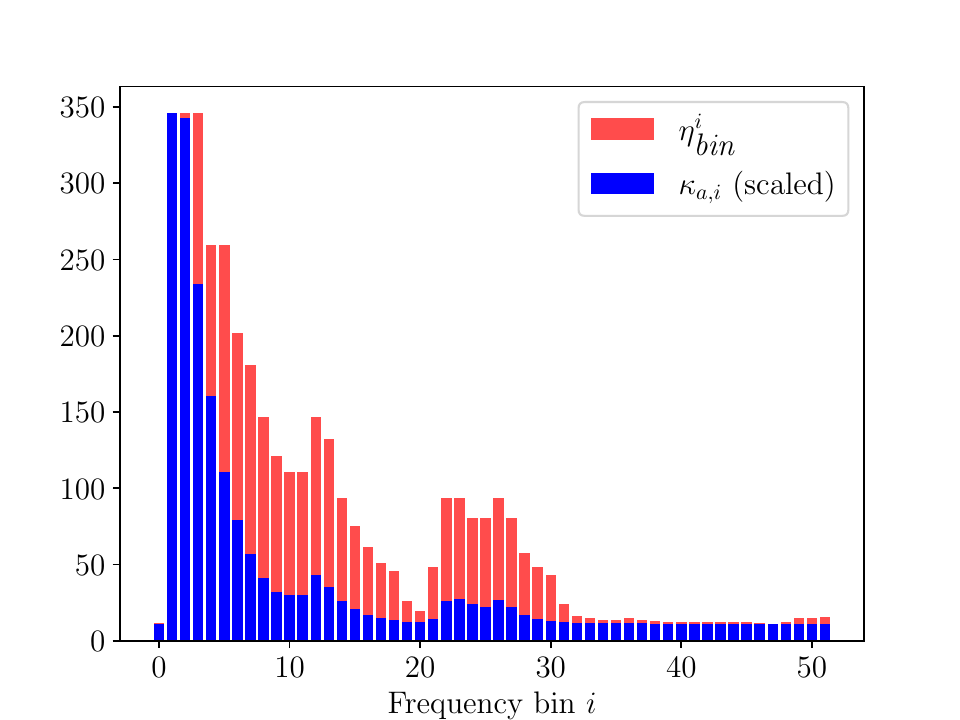}
    \caption{Bar plot of the compression ratio ($\mathcal{C}^i_{bin}$) evaluated for a tolerance of $10^{-9}$ and the scaled opacity $\kappa_{a,i}$ in each bin $i$. In some bins, the compression ratio exceeds the multigroup compression $\mathcal{C} = 144$. Nevertheless, the overall compression ratio across all bins is only about $58$, which is lower than $\mathcal{C}$ by a factor of $2.5$. Another noteworthy observation is that the binwise compression behavior reflects the underlying opacity structure: radiation tends to be more isotropic in optically thicker bins, leading to improved compressibility.}
    \label{fig:binwise_comp}
\end{figure}

From the above tests, we observe that our solution format requires significantly less storage than the traditional approach. However, an additional question arises: is it more efficient to run a TT-solver and store the solution for each frequency group independently, or to represent the solution across all groups as a single TT object?
The first option may seem advantageous when the solutions within certain groups are highly structured and low-rank, thereby reducing the total number of parameters. The second option, on the other hand, enables us to capture and exploit structure along the frequency domain. If the TT solution for bin $i$ has ranks $r^i_{x\theta}$ and $r^i_{\theta\phi}$, we define the number of bin-wise parameters as
\begin{align}
    N^i_{bin} := N_X r^i_{x\theta} + r^i_{x\theta} N_\theta r^i_{\theta\phi} + r^i_{\theta\phi} N_\phi \notag
\end{align}
and the bin-wise compression ratio
\begin{align}
     \mathcal{C}^i_{bin} := \frac{N_X N_\theta N_\phi}{N^i_{bin}}.\notag
\end{align}
We present the results for bin-wise compressions \revtext{for the original formulation of Graziani's problem (Section \ref{sec:Graziani_trad})} in Figure \ref{fig:binwise_comp} and Table \ref{tab:binwise_comp}. Figure \ref{fig:binwise_comp} presents a bar plot of $\mathcal{C}^i_{bin}$ for each frequency bin $i$. These values are obtained by truncating a low-tolerance TT solution in each bin independently at the final time step, using a tolerance of $10^{-9}$. For comparison, the multigroup TT solution—evaluated after truncating the same low-tolerance solution at the same final tolerance of $10^{-9}$—achieves a compression ratio of $\mathcal{C}=144$. It is observed that $\mathcal{C}^i_{bin}$ exceeds the multigroup compression ratios in certain bins. However, the aggregate number of parameters required to represent all bin-wise solutions ($= \sum_{i=1}^{N_f} N^i_{bin}$) remains substantially larger than that of the multigroup solution. We also compare the compression behavior with the opacity structure of the problem \revtext{(see Figure \ref{fig:graziani_result})} by plotting scaled opacities in each bin. We comment that the binwise compression pattern follows the opacity distribution. This due to the fact that in bins with larger optical depth, the radiation field becomes more isotropic and yields higher compression by exploiting low-rank structure with respect to the angular domain.

Table \ref{tab:binwise_comp} summarizes the compression achieved by the multigroup solutions relative to the bin-wise case across three different rounding tolerances---$10^{-3}$, $10^{-6}$, and $10^{-9}$. The results indicate that representing the full multigroup solution as a TT consistently yields superior compression performance compared to running a TT-solver in each bin independently for the one-spatial dimensional Graziani problem considered here. For multi-spatial dimensional problems where $N_X \gg N_\nu, N_\theta, \text{ and } N_\phi$, the storage for the extra angular cores required in the binwise approach will be relatively small compared to the storage required for the spatial cores.  As a result, the storage requirements for the two approaches may be closer, with the binwise approach possibly winning in some cases where the average $r^i_{x\theta}$ is less than $r_{z \theta}$.

\begin{deluxetable}{cc}
\label{tab:binwise_comp}
\tablecaption{Compressions obtained in the multigroup TT representation relative to the bin-wise case across various tolerances.}
\tablehead{
Tolerance &
$\left[\sum_{i=1}^{N_f} N^i_{bin} \right] / N_{TT}$
}
\startdata
        \hline
        $10^{-3}$ & $1.7$ \\
        $10^{-6}$ & $2$ \\
        $10^{-9}$ & $2.5$ \\
        \hline
\enddata
\end{deluxetable}

\subsection{Internal structure of the combined core}\label{subsubsec:core_structure}
\begin{figure}
    \centering
    \includegraphics[width=0.8\linewidth]{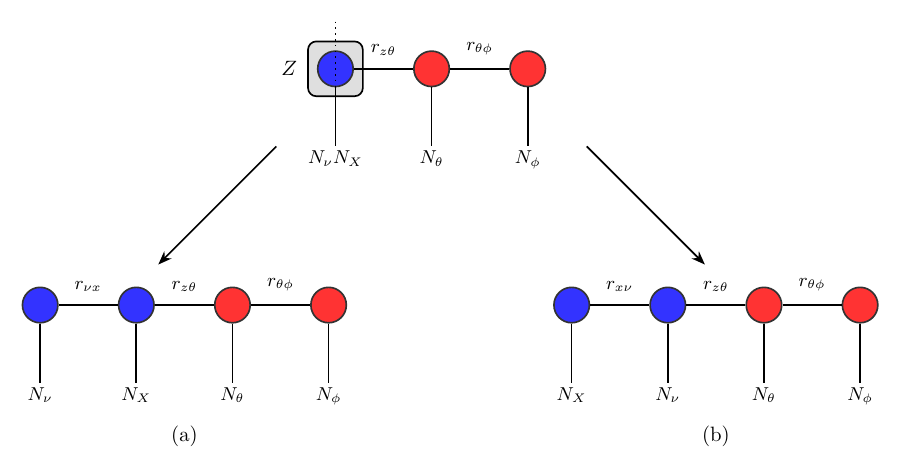}
    \caption{The figure illustrates two possible ways to split the spatio-spectral core $Z$ into separate TT cores for space and frequency. These choices lead to distinct TT topologies with different core orderings and corresponding definitions of the internal ranks. The preference for exploiting $\rone$ or $\ronex$ depends on both the discretization and the problem characteristics. For example, the multigroup hohlraum problem naturally favors topology (a), whereas problems such as stellar irradiation more effectively exploit low-rank structure under topology (b).}
    \label{fig:3d4dtt}
\end{figure}

We now examine results concerning the internal structure of the combined core $Z$ and discuss how a low internal rank may be exploited in theory to achieve further compression. We first introduce the notation and definitions used throughout this section. The internal $\delta$-rank of $Z$ is denoted by either $\rone$ or $\ronex$. Figure \ref{fig:3d4dtt} illustrates these internal ranks, which correspond to the TT ranks that would arise in two different hypothetical formulations with separate TT cores for the space and frequency dimensions (where $\rone$ denotes the rank of a hypothetical TT decomposition that leads with the frequency core, and $\ronex$ denotes the rank of a decomposition that begins with the spatial core). Throughout the remainder of this work, we explicitly distinguish between the TT ranks $\rtwo$, $\rthree$ and the internal ranks $\rone$ and $\ronex$.  One can obtain $\rone$ and $\ronex$ corresponding to a local error tolerance $\delta$ by performing a $\delta-\text{SVD}$ on $\texttt{reshape}(Z^m, [N_\nu, N_X \rtwo])$ and $\texttt{reshape}(Z^m, [N_X, N_\nu \rtwo])$ respectively. In both cases, $\Theta$ and $\Phi$ must be orthogonalized to ensure that the global error is equal to the local one.

Building on the preceding discussion, we introduce additional compression metrics, namely the relative compression ratios
\begin{align}
    \mathcal{C}_{rel}^\nu := \frac{N_{TT}}{N_\nu \rone + N_X \rone \rtwo + N_\theta \rtwo \rthree + N_\phi \rthree}\label{eq:rel_compression_nu}
\end{align}
and
\begin{align}
    \mathcal{C}_{rel}^x := \frac{N_{TT}}{N_X \ronex + N_\nu \ronex \rtwo + N_\theta \rtwo \rthree + N_\phi \rthree},\label{eq:rel_compression_x}
\end{align}
that quantify the \textit{additional} compression benefits that can be obtained by having separate cores for space and frequency.\footnote{We emphasize again that the extra compression is practically attainable only when the transport and source operators can be applied independently to the space and frequency cores.} $\mathcal{C}_{rel}^\nu$ is the additional compression obtained by exploiting $\rone$ whereas $\mathcal{C}_{rel}^x$ is the additional compression obtained by exploiting $\ronex$. Comparing $\mathcal{C}_{rel}^\nu$ and $\mathcal{C}_{rel}^x$ allows us to determine the better TT representation between (a) and (b) in Figure \ref{fig:3d4dtt} from a compression point of view.  

Knowing a priori which TT topology offers superior compressions is often nontrivial.  \cite{einkemmer24} argues that placing ``less separable" dimensions (i.e., modes connected by a higher rank) closer together in the TT representation (see Figure \ref{fig:3d4dtt}) tends to improve compression performance.  For example, in the optically thin regime (i.e., in the $\sim$free-streaming limit), the transport operator promotes strong coupling between space and angle potentially leading to limited separability between those dimensions.  Problems in this regime, therefore, may benefit from a TT topology that positions the spatial and angular cores ``closer" to one another (i.e., adopting internal rank $\rone$).  Competing, however, is separability between frequency and angle---but in this work, we adopt opacities $\kappa_a$ and $\kappa_s$ that are \textit{not} a function of angle, likely promoting high separability between these dimensions. Even under these considerations, the optimal topology is still nuanced as the TT solution size is dependent on problem discretization.  For problems where $N_x \gg N_\nu, N_\theta, N_\phi$, topology \ref{fig:3d4dtt}b might be advantageous over \ref{fig:3d4dtt}a as the (likely) dominant $N_X \ronex$ term in the denominator of Equation (\ref{eq:rel_compression_x}) might be significantly smaller than the dominant $N_x \rone \rtwo$ term in the denominator of Equation (\ref{eq:rel_compression_nu}).

We take the opportunity to summarize the solver performance for all test problems by reporting $\mathcal{C}$ and $\mathcal{S}$, and quantify the compression gains arising from the internal structure of $Z$ through $\mathcal{C}_{rel}$ in Table \ref{tab:conclusion}. Below, we discuss some results on the internal ranks $\rone$ and $\ronex$ across all test cases.

\paragraph{\textbf{Hohlraum and Thermal Relaxation}} Irrespective of the boundary condition in the multigroup hohlraum problem or the opacity structure in the thermal relaxation test, the transport physics for each frequency group is independent and results in a solution that is perfectly compressible in frequency. Put differently, for all times one can represent the solution as as $k_f \cdot g(x, \theta, \phi)$ where $k_f$ is a frequency-dependent scalar. This implies that $\rone=1$ and $Z$ has high internal structure hence promoting high $\mathcal{C}_{rel}^\nu$.  The multigroup hohlraum test exhibits $\mathcal{C}_{rel}^\nu =4.9$, meaning that a specialized solver that adopts topology in Figure \ref{fig:3d4dtt}a would permit $4.9\times$ additional compression.  Albeit we see $\mathcal{C}_{rel}^\nu = \mathcal{C}_{rel}^x \approx 1$ in the thermal relaxation problem, we note that this problem is effectively purely local and does not require spatial discretization. It is instructive to compare the increase in storage cost for conventional and TT representations when transitioning from a gray to a multigroup formulation in problems that are rank-1 in frequency. In such cases, the storage required by a traditional representation grows linearly with the number of frequency groups $N_\nu$, whereas the TT representation in Figure \ref{fig:3d4dtt}a requires only $N_\nu$ additional parameters beyond those needed for the gray solution. This behavior is intuitive for the hohlraum problem, where each frequency group mirrors the gray dynamics and differs only by a group-dependent scaling factor, which is efficiently captured by the additional $N_\nu$ parameters in the multigroup TT solution. Finally, we note that although the Hohlraum solution does not admit $\ronex = 1$, having $\rone = 1$ substantially constrains the maximum attainable $\ronex$. As a result, topology \ref{fig:3d4dtt}b yields compressions that are comparable but smaller $(\mathcal{C}_{rel}^x = 4.8)$ than those achieved with topology \ref{fig:3d4dtt}a.

\begin{figure*}
    \centering
    \includegraphics[width=1.0\linewidth]{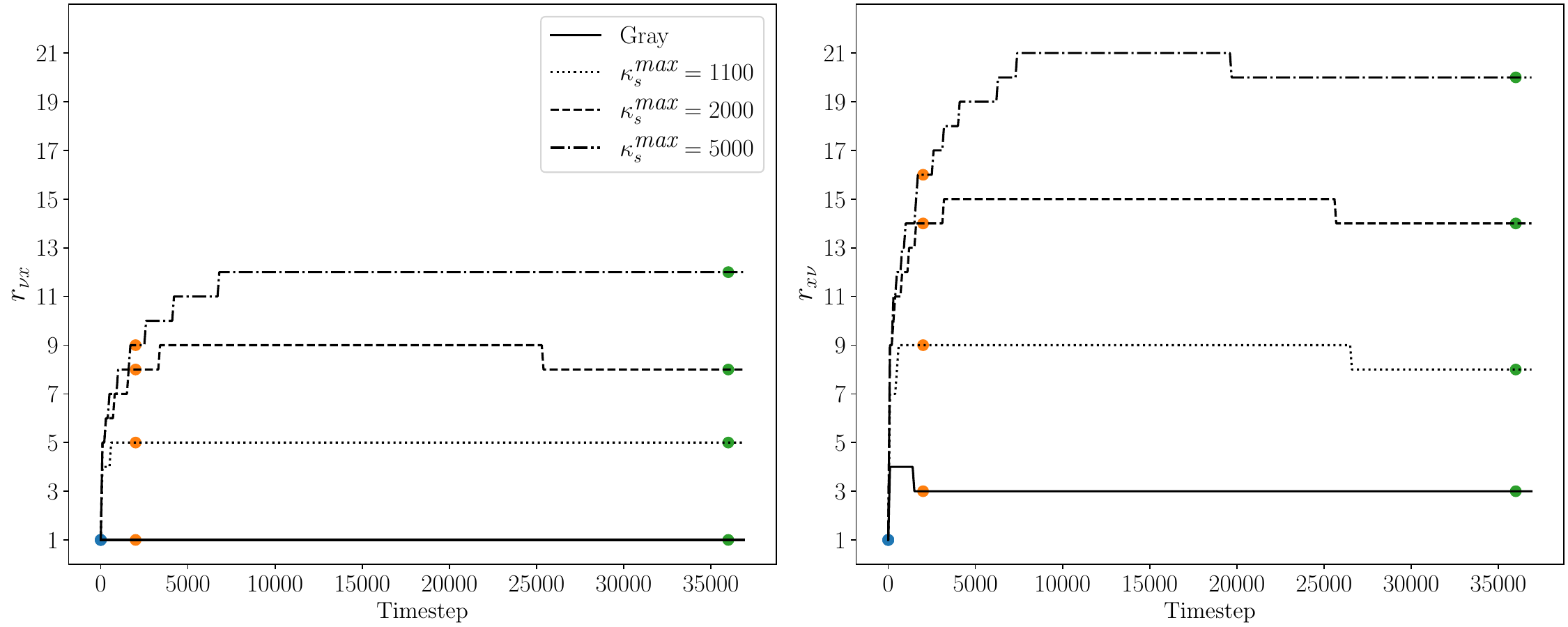}
    \caption{Evolution of the internal rank $\rone$ (left) and $\ronex$ (right) for progressively larger ranges of $\kappa_{s,f}$. Colored markers indicate the ranks at time steps $1$, $2000$, and $38000$ with the corresponding solutions at these times shown in Figure \ref{fig:diffusion}. The internal ranks grows from its initial value of 1 as the groups begin to diffuse at different rates, reaches a peak, and subsequently decreases as the solution decays toward zero. We emphasize that a wider spread in opacities leads to greater dissimilarity among the groupwise solutions, necessitating higher TT ranks for accurate representation. This trend is reflected in the growth of the maximum value achieved by $\rone$ and $\ronex$ as the opacity range is increased.}
    \label{fig:split_diffusion_ranks}
\end{figure*}

\paragraph{\textbf{Gaussian Diffusion}} We next investigate the evolution of  $\rone$ and $\ronex$ for the Gaussian diffusion problem and examine its dependence on the range of scattering opacities employed in the problem. In this problem, the radiation field approaches isotropy, and variations in the internal ranks can therefore be attributed to the spatio-spectral complexity arising from frequency-dependent scattering opacities.  The effects of scattering opacity range are observed in Figure \ref{fig:split_diffusion_ranks} where we plot the evolution of the internal ranks for different values of $\kappa_s^{max}$. The initial solution is identical for each group and is therefore a rank-1 TT. The ranks increase as the solutions begin to diverge inside different groups at varying rates due to different values of $\kappa_{s,f}$. The radiation energy asymptotically approaches zero across all frequency groups, leading to a decrease in $\rone$ and $\ronex$. We also remark that the internal ranks attain a greater maximum value when the span of opacities present in the problem is large. A wider range yields less similar solution profiles across different groups and therefore leads lower internal structure within $Z$. The observed values of $\rone$ and $\ronex$ remain low despite the use of randomly sampled opacities, indicating structure within $Z$. In general, we find that $\mathcal{C}_{rel}^\nu$ is greater than $\mathcal{C}_{rel}^x$. The difference, however, is small (i.e., both lie in the range of $2$ to $3$).  Due to global $\sim$isotropy in this problem, separability in space–angle and frequency–angle is unlikely to provide an advantage for a TT topology with internal rank $\rone$.  Alongside the above considerations, a discretization with $N_X \simeq N_\nu$ suggests little difference in compression performance between the two topologies, consistent with our observations.

\begin{figure}
    \centering
    \includegraphics[width=1.0\linewidth]{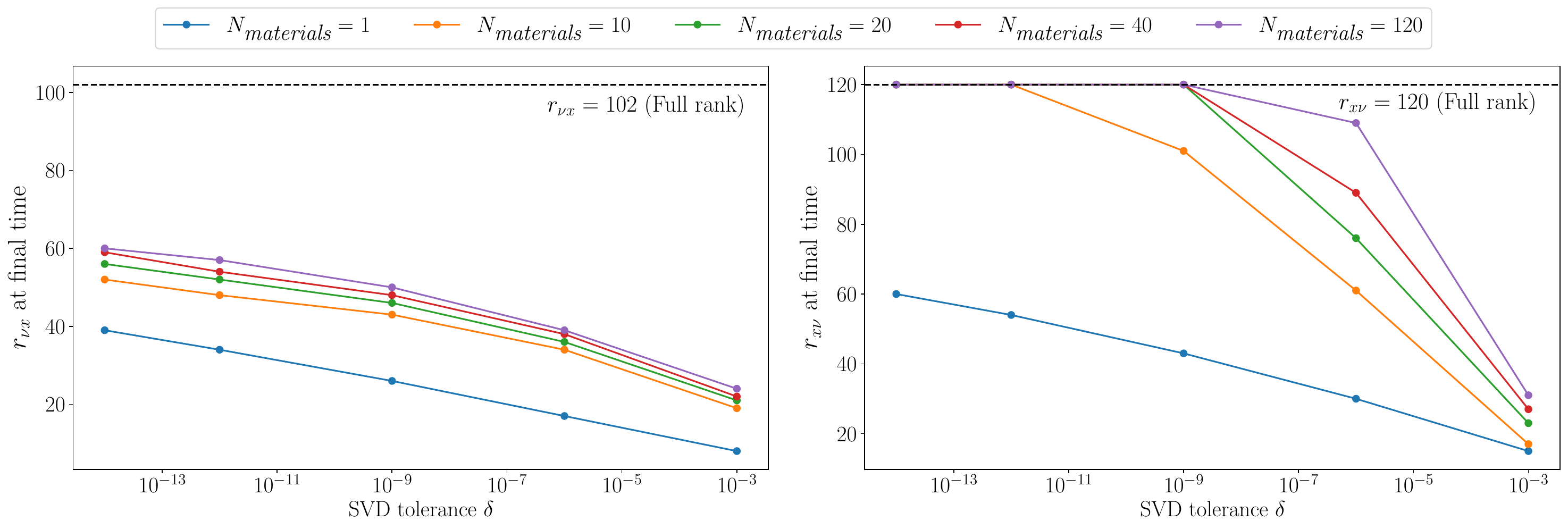}
    \caption{Evolution of the internal ranks (left) and their dependence on the SVD tolerance $\delta$ (right) for test cases with 1, 5, 10, 20, 40, and 100 spatially distributed materials and a rounding tolerance of $10^{-14}$. The internal ranks grow with increasing $N_{materials}$ and tighter $\delta$. We observe that $\rone$ always remains well below its full rank, demonstrating low-rank internal structure within $Z$  despite the problem’s complexity. On the other hand, $\ronex$ approaches its full rank for lower values of $\delta$. This is reflected in the relative compression numbers which tend to be higher for TT-topology in Figure \ref{fig:3d4dtt}a compared to the one in Figure \ref{fig:3d4dtt}b.}
    \label{fig:comp_graziani_vary}
\end{figure}

\paragraph{\textbf{Graziani Prompt Spectrum}} We assess the internal rank behavior for Graziani's prompt spectrum problem with spatially varying opacities where the number of materials in the problem is equal to the matrix-rank of the absorption opacity field, i.e., $\mathrm{rank}(\rho \kappa_{a,f}) = N_{materials}$ with $\rho \kappa_{a,f} \in \mathbb{R}^{N_X \times N_\nu}$. Since $\mathrm{rank}(\rho \kappa_{a,f})$ directly influences the growth of the internal ranks, we anticipate that $N_{materials}$ will affect the internal rank behavior. Figure \ref{fig:comp_graziani_vary} summarizes these results, showing the evolution of the internal ranks in the left plots and their variation with respect to $\delta$ on the right, for six test cases with 1, 5, 10, 20, 40, and 100 materials distributed in space. As expected, these ranks increase with larger $N_{materials}$ and tighter tolerances $\delta$. For high material counts and small $\delta$, $\rone$ remains well below its full rank $\min(N_\nu, N_X N_\theta N_\phi)$ whereas $\ronex$ reaches its full rank $\min(N_X, N_\nu N_\theta N_\phi)$. The stronger low-rank structure in the former yields larger relative compression ratios, with $\mathcal{C}_{rel}^\nu$ decreasing from $2.9$ at $N_{materials} = 1$ to $2$ at $N_{materials} = 100$, compared to $\mathcal{C}_{rel}^x$, which drops from $2.1$ to $1$ over the same range. In the Graziani prompt spectrum test, more pronounced deviations away from global isotropy disfavor separability in space and angle, thereby encouraging a TT topology with $\rone$ (reflected in $\mathcal{C}_{rel}^x<\mathcal{C}_{rel}^\nu$).  We conclude by noting that, under the current tensor decomposition, the TT ranks remain largely insensitive to opacity-driven complexity, whereas the internal ranks reflect these effects. Furthermore, for problems exhibiting strong coupling between space and frequency, splitting the spatio-spectral core offers little to no additional benefit.
\paragraph{\textbf{Stellar Irradiation}} The stellar irradiation problem, which demands fine spatial and angular resolution and would ordinarily incur prohibitive storage costs, benefits substantially from the TT representation, as evidenced by the high $\mathcal{C}$ ($= 4583$). Outside the central clump, the optically thin background promotes strong space-angle coupling (i.e., in the $\sim$free-streaming limit), thereby favoring a TT topology with $\rone$. Nevertheless, we observe $\mathcal{C}_{rel}^\nu$ $=2$ and $\mathcal{C}_{rel}^x$ $=7$. We hypothesize that the higher $\mathcal{C}_{\text{rel}}^x$ is likely due to a combination of spatial symmetry in the solution about the $45^\circ$ diagonal and the fact that $N_X \gg N_\nu$. Consequently, the dominant storage cost in both topologies scales with $N_X$; however, in topology \ref{fig:3d4dtt}a this term is weighted by only one rank 
$\ronex$, yielding superior compression.

\begin{deluxetable}{lccccccccccccc}
\tablecaption{Summary of results for all test cases presented in Section \ref{sec:tests}. The table reports solver settings, including the rounding method, rounding tolerance and SVD tolerance $\delta$, along with performance metrics such as the minimum compression ratios, defined in (\ref{eq:compression}), (\ref{eq:rel_compression_nu}), and (\ref{eq:rel_compression_x}), over the course of the simulation as well as the achieved speedup $\mathcal{S}$. \revtext{Note that the compression metrics are evaluated after truncation of the TT-solution at the end of a timestep.} The file sizes of the solutions at $(\mathcal{C})^{min}$ are also reported. The consistently high total and relative compression ratios demonstrate that, in addition to using a TT representation, employing a separate frequency core is advantageous in most cases. The dashes in the Hohlraum and thermal relaxation results signify insensitivity to tolerance value. \label{tab:conclusion}}
\tablehead{
Test &
Case &
$N_X$ &
$N_\nu$ &
$N_\theta$ &
$N_\phi$ &
Rounding method &
Rounding tol &
SVD tol $\delta$ &
$(\mathcal{C})^{min}$ &
$(\mathcal{C}_{rel}^\nu)^{min}$ & 
$(\mathcal{C}_{rel}^x)^{min}$ & 
$\mathcal{S}$ & 
File size
}
\startdata
Hohlraum 2d
& $I_{f,\mathrm{init}}=\varepsilon_f(T_{\mathrm{init}})$
& $128^2$ & $10$ & $512$ & $1024$ 
& \revtext{SVD} & $10^{-4}$ & -- 
& \revtext{$4565$} & \revtext{$4.8$} & \revtext{$4.7$} & $60$ & \revtext{$151$} MB \\
Hohlraum 2d
& $I_{f,\mathrm{init}}=1/N_\nu$
& $128^2$ & $10$ & $512$ & $1024$ 
& \revtext{SVD} & $10^{-4}$ & -- 
& \revtext{$4565$} & \revtext{$4.8$} & \revtext{$4.7$} & $60$ & \revtext{$151$} MB \\
\hline
Thermal relaxation
& $N_\nu = 3$
& $4$ & $3$ & $512$ & $1024$ 
& \revtext{SVD} & -- & -- 
& $4000$ & $1$ & $1$ & $0.8$ & $14$ KB \\
Thermal relaxation
& $N_\nu = 100$
& $4$ & $100$ & $512$ & $1024$ 
& \revtext{SVD} & -- & -- 
& $10^5$ & $1.18$ & $1.18$ & $25$ & $21$ KB \\
\hline
Gaussian diffusion
& $\kappa_s^{\max} = 1000$
& $128$ & $100$ & $1024$ & $2048$ 
& \revtext{SVD} & $10^{-8}$ & $10^{-12}$ 
& $3.5 \times 10^5$ & $3$ & $2.8$ & $5865$ & $916$ KB \\
Gaussian diffusion
& $\kappa_s^{\max} = 1100$
& $128$ & $100$ & $1024$ & $2048$ 
& \revtext{SVD} & $10^{-8}$ & $10^{-12}$ 
& $3.5 \times 10^5$ & $2.7$ & $2.6$ & $5865$ & $916$ KB \\
Gaussian diffusion
& $\kappa_s^{\max} = 2000$
& $128$ & $100$ & $1024$ & $2048$ 
& \revtext{SVD} & $10^{-8}$ & $10^{-12}$ 
& $3.5 \times 10^5$ & $2.6$ & $2.5$ & $5865$ & $916$ KB \\
Gaussian diffusion
& $\kappa_s^{\max} = 5000$
& $128$ & $100$ & $1024$ & $2048$ 
& \revtext{SVD} & $10^{-8}$ & $10^{-12}$ 
& $3.5 \times 10^5$ & $2.5$ & $2.2$ & $5865$ & $916$ KB \\
\hline
Graziani
& \cite{graziani2008computational}
& $120$ & $52$ & $64$ & $128$ 
& \revtext{SVD} & $10^{-6}$ & $10^{-9}$ 
& $296$ & $2$ & $2.1$ & $1.5$ & $1.5$ MB \\
Graziani
& $N_{\mathrm{materials}} = 1$
& $120$ & $102$ & $64$ & $128$ 
& \revtext{SVD} & $10^{-6}$ & $10^{-9}$ 
& $308$ & $3$ & $2.3$ & $2$ & $2.8$ MB \\
Graziani
& $N_{\mathrm{materials}} = 10$
& $120$ & $102$ & $64$ & $128$ 
& \revtext{SVD} & $10^{-6}$ & $10^{-9}$ 
& $308$ & $2$ & $1.1$ & $2$ & $2.6$ MB \\
Graziani
& $N_{\mathrm{materials}} = 20$
& $120$ & $102$ & $64$ & $128$ 
& \revtext{SVD} & $10^{-6}$ & $10^{-9}$ 
& $308$ & $1.9$ & $1$ & $2$ & $2.6$ MB \\
Graziani
& $N_{\mathrm{materials}} = 40$
& $120$ & $102$ & $64$ & $128$ 
& \revtext{SVD} & $10^{-6}$ & $10^{-9}$ 
& $308$ & $1.9$ & $1$ & $2$ & $2.5$ MB \\
Graziani
& $N_{\mathrm{materials}} = 120$
& $120$ & $102$ & $64$ & $128$ 
& \revtext{SVD} & $10^{-6}$ & $10^{-9}$ 
& $308$ & $1.8$ & $1$ & $2$ & $2.5$ MB \\
\hline
Stellar irradiation
& --
& $310^2$ & $20$ & $512$ & $1024$ 
& \revtext{Gram} & $10^{-3}$ & $10^{-3}$ 
& $4583$ & $2$ & $7$ & $232$ & $1.8$ GB \\
\enddata
\end{deluxetable}
\subsection{Pointwise errors}
\revtext{In this subsection, we use a rounding tolerance of $10^{-6}$ within the step-and-truncate procedure for Graziani’s multi-material problem, where the solution $\hat{I} = Z\Theta\Phi$ is rounded at every timestep. After the final timestep, we post-process the rounded solution by applying a $\delta-$truncated SVD to the core $Z$. This truncation provides a global approximation of the solution while ensuring that the relative Frobenius norm error remains below $\delta$.}\footnote{In a hypothetical implementation that splits the combined core $Z$, the SVD tolerance $\delta$ would be equivalent to the rounding tolerance used to truncate $\rone$/$\ronex$.} This approximation roughly has the effect of limiting the accuracy of the solution to a window extending from the maximum value down to approximately the product of the tolerance and the maximum value. Outside this range, pointwise errors may be incurred up to the relative error. We quantify these pointwise errors by introducing the error field $\gamma(\delta)$ which is the relative error between the mean intensity from the TT solver \revtext{after the final timestep and the mean intensity when the internal rank of $Z$ in the final solution is restricted by performing the $\delta-$SVD procedure.}

\begin{figure}
    \centering
    \includegraphics[width=0.45\linewidth]{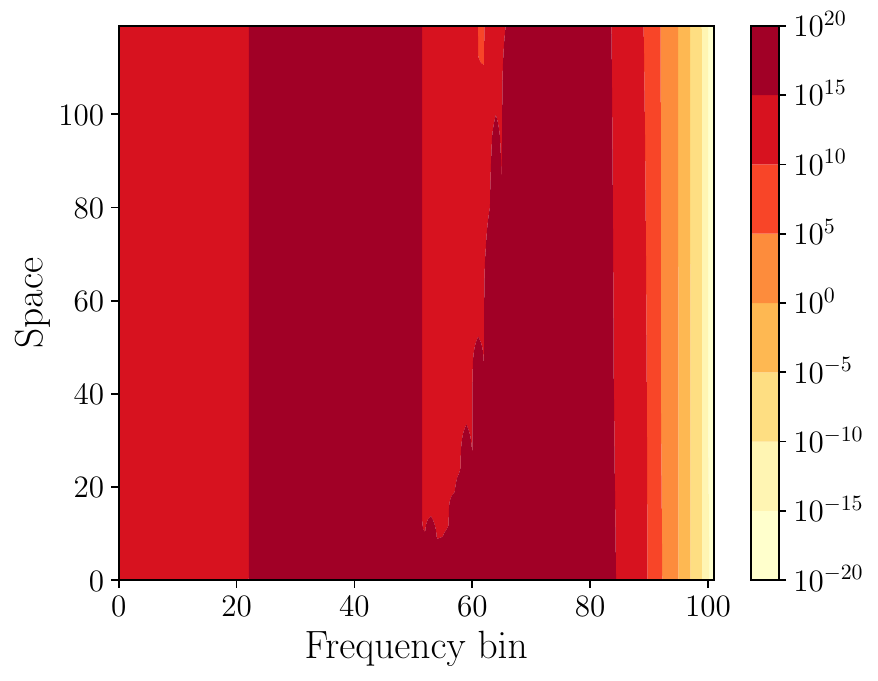}
    \caption{Mean intensity field produced by our solver for the 10-material case. This captures all the spatio-spectral interactions as opposed to the results shown in Figure \ref{fig:tol_effect_graziani_vary}a, where the internal rank of $Z$ is limited to a tolerance $\delta$.}
    \label{fig:J_merged}
\end{figure}

\begin{figure*}
\centering

\gridline{
    \fig{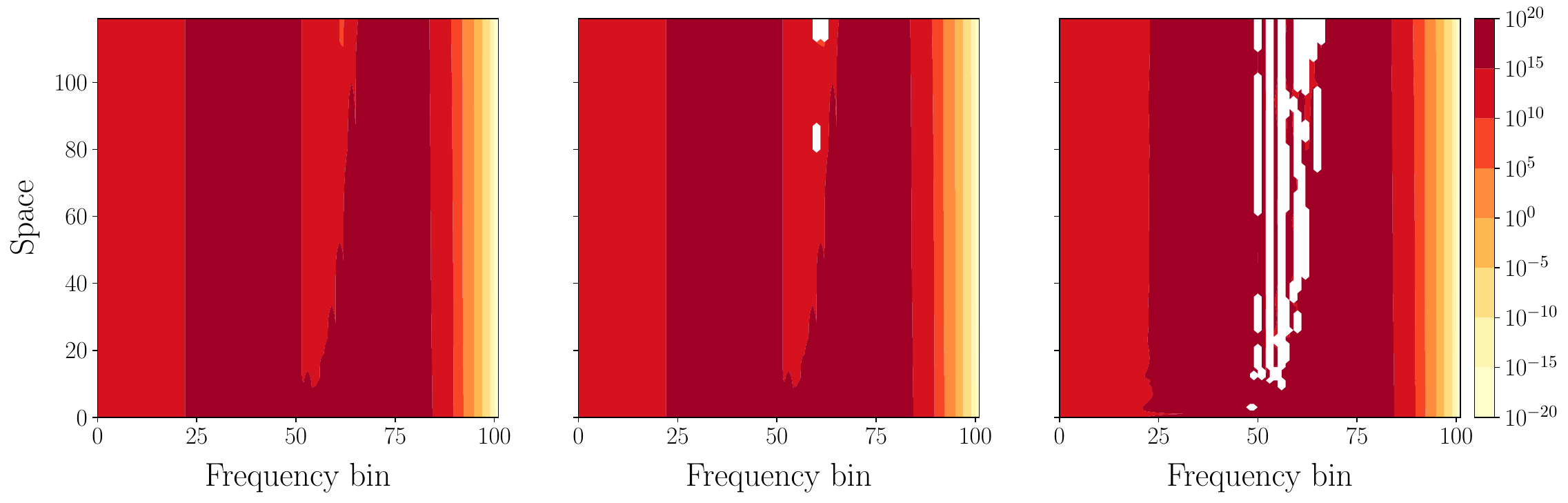}{0.90\textwidth}{(a) Mean intensity field $J$.}
}

\gridline{
    \fig{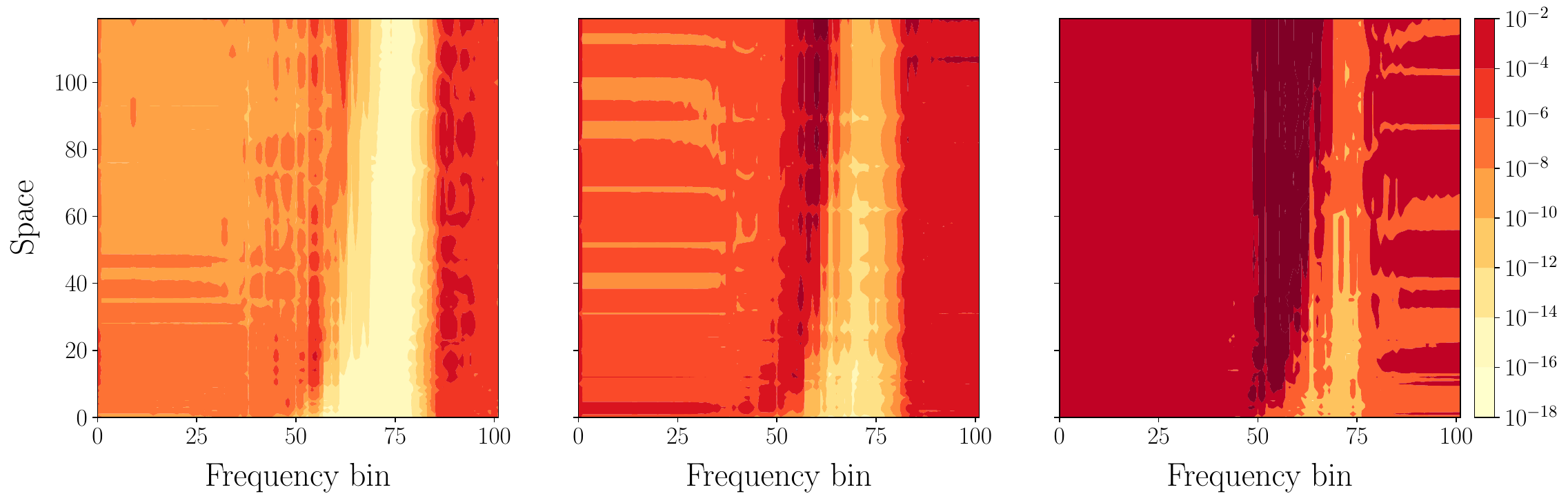}{0.90\textwidth}{(b) Absolute relative error.}
}

\gridline{
    \fig{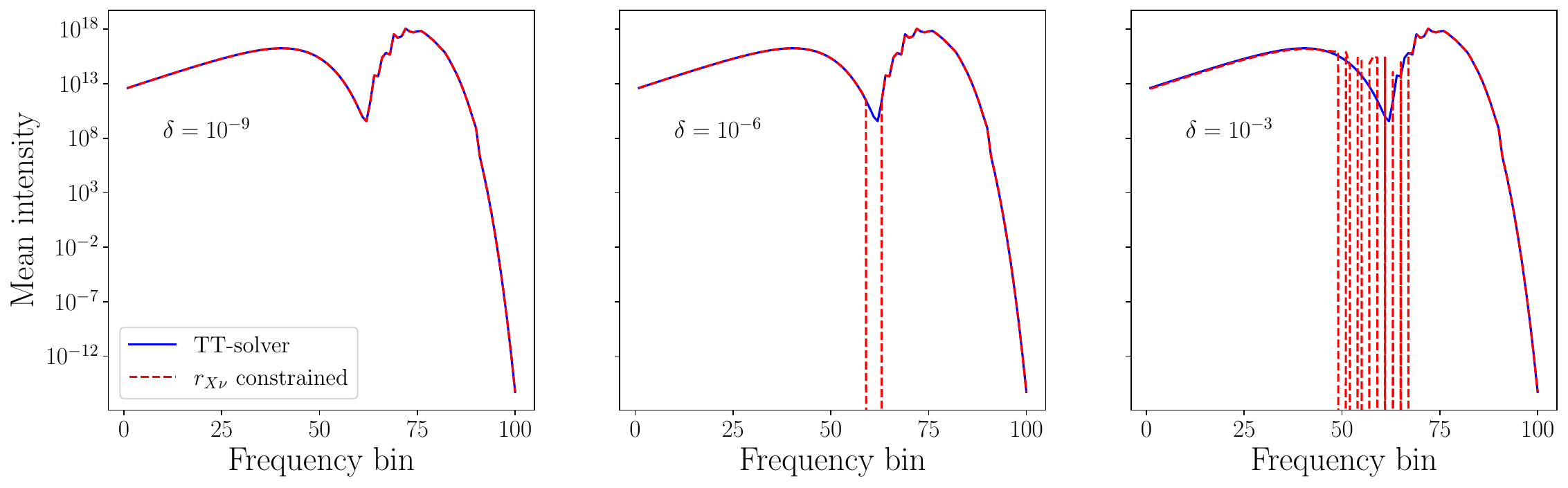}{0.85\textwidth}{(c) Cross-section plot of $J$ at a fixed spatial point.}
}
\caption{
Mean intensity and error plots for the 10-material case when the internal rank $\rone$ is constrained using $\delta$. 
The left, center, and right plots correspond to $\delta=10^{-9}$, $10^{-6}$, and $10^{-3}$, respectively.  
White regions in the contour plots indicate negative values, which appear as small oscillations in the cross-section plots. These errors diminish as $\delta$ decreases.
}
\label{fig:tol_effect_graziani_vary}
\end{figure*}

The effect of these pointwise errors can be observed in Graziani’s problem, where initializing the specific intensity $I$ with a Planck distribution produces a broad range of values in $I$. Figure \ref{fig:J_merged} shows the mean intensity from our solver and sub-figures \ref{fig:tol_effect_graziani_vary}a, \ref{fig:tol_effect_graziani_vary}b, and \ref{fig:tol_effect_graziani_vary}c display the mean intensity field, error field $\gamma(\delta)$, and mean intensity at a fixed spatial point respectively when the internal rank $\rone$ is constrained by $\delta$. The results are obtained for $\delta=$ $10^{-9}$, $10^{-6}$, and $10^{-3}$ while the TT-rounding tolerance is fixed at $10^{-6}$ for all cases. These results highlight how the value of $\delta$ can constrain the dynamic range of the solution that is accurately captured. For instance, in the $\delta = 10^{-6}$ case (last row), values outside the interval from $10^{18}$ (the maximum) down to $10^{18-6} = 10^{12}$ show numerical artifacting: white patches in column 1 correspond to negative values, which appear as oscillations in the log-log plot of column 3. Furthermore, the error surpasses 1 outside this range, as seen in column 2. The reduction in observed artifacts as $\delta$ decreases indicates that tighter tolerances provide a more accurate solution over a larger dynamic range. While we present results only for $\rone$, similar patterns are observed for $\ronex$.

The pointwise errors in the mean intensity are strongly dependent on $\delta$, highlighting a tradeoff between increased compression and the fidelity of the mean intensity field. Nonetheless, a larger $\delta$ may be acceptable if the application does not demand resolving such a wide dynamic range. We end this discussion by reiterating that these artifacts do not occur in our solver and appear only when the internal $\delta$-rank of $Z$ is leveraged by splitting it into separate space and frequency cores.

\subsection{Trade-offs and Implications}
Despite the attractive potential for additional compression afforded by splitting $Z$, we choose to not pursue this approach in the present work due to practical feasibility considerations discussed earlier---that is, several key quantities, such as opacities and emissivities, depend simultaneously on space and frequency, and these quantities must have the same format as the TT solution in order to interact through standard TT operations. If the TT solution were to employ separate spatial and frequency cores, the spatio-spectral quantities would first need to be split accordingly before being added to or multiplied with the solution. This process introduces additional singular value decompositions, along with the associated truncation errors and computational overhead. Such costs can become prohibitive, particularly if SVDs must be performed at every time step. 

Besides implementation concerns, there are other potential issues which preclude us from splitting $Z$. For scenarios exhibiting significant spatio-spectral complexity, such as those arising from intricate opacity structures, the additional compression advantage diminishes. Furthermore, pointwise errors are observed in the radiation energy profiles for the fully low-rank case. This behavior arises because Planckian distributions can introduce large variations in magnitude depending on the discretization of the frequency domain—specifically, the integrated Planckian values within bins can vary substantially based on their proximity to the spectral peak. Although this effect is generally negligible, in highly complex setups (e.g., Graziani’s problem with spatially varying opacities), the finite precision of singular values in SVD-based operations such as rounding can lead to localized discrepancies. Nonetheless, the relative Frobenius norm of such errors remains bounded below a user-defined tolerance, ensuring controlled numerical accuracy.

\section{Conclusion}\label{sec:conclusion}
In this work, we have presented an algorithm to solve the multigroup radiation transport equation using a TT-solver. Our solver employed a TT decomposition that contained a single ``merged" TT core for the spatial and frequency dimension.  We tested our solver on a variety of test problems and analyzed its performance.

High compression ratios were consistently achieved with the TT solver across all test cases. The gray versions of the hohlraum, thermal relaxation ($N_\nu = 100$), and Gaussian diffusion problems covered in \cite{gorodetsky25} achieve compressions of approximately $3 \times 10^3$, $1300$, and $1.7 \times 10^4$, respectively, relative to their traditional gray representations. In this work, we observe even larger compressions when applying the multigroup TT solver to these problems' frequency-dependent counterparts at the same discretization and rounding tolerance. The reason is that the multigroup TT solution exploits correlations not only between spatial and angular dimensions, as in the gray case, but also between frequency and angular dimensions, thereby maintaining relatively low TT ranks. While the multigroup TT representation does require more parameters than its gray counterpart, this increase is lower compared to the traditional storage cost, which scales linearly with $N_\nu$.

\revtext{We observe speedups across all test problems. However, in contrast to the compression ratios, the observed speedups are more modest, which may necessitate parallelization for certain practical applications. Parallelization can be pursued along two primary directions. The first involves parallelizing computationally intensive TT operations—such as rounding \citep{al2022parallel}—which constitute a major bottleneck in the solver. The second adopts a domain decomposition strategy, wherein the spatial domain is partitioned into multiple subdomains that are distributed across processes. Within this framework, the solution is advanced in time in each subdomain, with the subdomain configuration dynamically updated at each timestep. This approach introduces several technical challenges, including load balancing and inter-process communication via MPI. We note that these components can be implemented within the TT framework, and work is currently underway to develop a block adaptive mesh refinement (AMR) TTTT solver.}

In addition, we also examined the low-rank structure within the combined space–frequency core $Z$ as a potential source of further compression. We introduced two different definitions of the internal rank ($\rone$ and $\ronex$) and linked them to alternative TT core orderings in which space and frequency are represented by separate cores (Figure \ref{fig:3d4dtt}). We studied their behavior in all test problems and compared the additional compression obtained by exploiting either of these internal ranks. For the hohlraum, Gaussian diffusion, and Graziani problems, topology \ref{fig:3d4dtt}a proved more efficient, whereas the stellar irradiation problem was better represented by topology \ref{fig:3d4dtt}b. We note that while additional compression is possible, its effectiveness depends on both the feasibility of applying transport and source operators in low-rank form and the underlying problem complexity. In cases with strong spatio-spectral coupling, such as complex opacity fields, the gains diminish, and small pointwise errors may appear due to SVD truncation effects.

Several aspects of the problem remain unexplored in this study. We briefly outline some of these limitations and potential directions for future work. First, no systematic strategy was used to partition the frequency domain into groups. The specific choice of bins, together with the associated radiation dynamics, directly influences the TT ranks, but the impact of this choice has not been examined here.

Second, we have only partially explored the topic of alternate TT-core orderings in our internal rank study. We concede that there are many possibilities that have not been investigated in this work. For example, a TT solution that adopts a $Z \Phi \Theta$ ordering as opposed to our 
$Z \Theta \Phi$ ordering might offer improved compression. We remark that identifying the optimal core ordering, or more broadly, the optimal tensor-network topology for high-dimensional data remains a challenging and open problem in the tensor-network literature.

\revtext{Third, we observe that the Gram rounding algorithm is unstable for Graziani’s problem. As discussed earlier, this problem introduces a wide range of values in the solution, leading to large condition numbers in the matrices involved in the rounding step. This issue is further amplified in Gram rounding, where forming Gram matrices effectively squares the condition number, resulting in well-known numerical stability challenges \citep{al2022parallel}. Therefore, we recommend using alternative rounding algorithms for such cases, while leaving the investigation of strategies to improve the robustness of Gram rounding to future work.}

Finally, we acknowledge that our solver does not take advantage of the low-rank structure of $Z$, owing to the absence of a feasible approach for applying the HLL transport and source operators to a TT solution with separated space and frequency dimensions. Potential remedies include randomized SVD techniques or cross-approximation–based methods, which we leave for future investigation.

\begin{acknowledgments}
A.D and A.A.G were supported by the Los Alamos National Laboratory (LANL) under the project ``Algorithm/Software/Hardware Co-design for High Energy Density applications'' at
the University of Michigan. A.D. was also supported, in part, by AFOSR Computational Mathematics Program under the Award \#FA9550-24-1-0246 to study the performance of randomized rounding in the MRT context.
P.D.M, J.C.D, C.D.M., J.M.M., and L.F.R. were supported by LANL under the M$^3$AP project. LANL is operated by Triad National Security, LLC, for the National Nuclear Security Administration of U.S. Department of Energy (Contract No. 89233218CNA000001). This document is approved for unlimited release under LA-UR-26-20665.
\end{acknowledgments}

\bibliography{refs}{}
\bibliographystyle{aasjournalv7}

\clearpage
\appendix

\begin{revblock}
\section{Algorithm cost}
\label{sec:appendix}

In this section, we analyze a single timestep of the algorithm to identify the dominant cost computational operation and the point of peak memory usage. The transport update procedure is illustrated in Figure \ref{fig:trans_op}.
The TT with red nodes denotes the solution at the beginning of the timestep. The orange TT represents the solution after padding the spatial core with ghost-cell values \citep[used in applying physical boundary conditions; see][for details]{gorodetsky25}. The blue TT corresponds to a rank-1 Dirichlet tensor containing boundary values in the ghost cells and zeros elsewhere. The green TT represents the solution after the transport update. Here, $N_{Xg} = N_X + N_g$ and $N_g$ denotes the number of ghost cells.
The transport operator is expressed as a sum of $n$ rank-1 TT operators and is applied to the sum of the ghost-cell-padded solution and the Dirichlet boundary-condition TT
where the value of $n$ depends on the number of spatial dimensions $d_X$. More precisely, $n = 2 d_X$ for all three numerical fluxes considered in this work.
Equivalently, this operation can be viewed as applying each operator separately to both the solution and the boundary-condition TT.
In practice, this results in the creation of $n$ additional copies of the original solution (red), each padded with ghost cells to form the orange representation. Applying a single rank-1 transport operator $i$ to the padded solution produces an updated tensor $TT^i$ that retains the same TT ranks $\rtwo,\rthree$ as the original solution. Likewise, the boundary-condition TT remains rank-1 after application of operator $i$. The $n$ updated solution tensors $TT^i$ are stored separately, while the $n$ updated boundary-condition tensors are summed, resulting in a TT whose two ranks are both equal to $n$. Consequently, after the transport step and before rounding, the solution consists of the sum of:
(1) the initial solution with ranks $\rtwo,\rthree$,
(2) $n$ updated solution tensors, each with ranks $\rtwo,\rthree$, and
(3) the summed boundary-condition tensor with both TT ranks equal to $n$. Under this setup, one could explicitly form the sum of all intermediate TTs, represented by the graph with green nodes, which produces a TT with ranks $r_1=(n+1) \cdot \rtwo + n$ and $r_2=(n+1) \cdot \rthree + n$, or store each TT operand in the sum separately. The resulting tensor is then rounded, and the preferred storage strategy depends on the specific rounding algorithm being used. The transport step is followed by the source term update. The source term typically has a limited effect on increasing the TT ranks, as it raises them by only a value of 2 before rounding is applied. In contrast, the post-transport solution prior to rounding and the rounding operation itself, often create a storage and computational bottleneck within each timestep of our solver. For this reason, we focus on this stage in the later discussion.

\paragraph{Value of n}

A transport update is given by an application of a linear operator
\begin{equation*}
    I^{*} = \left(\mathcal{I} + \sum_{d=\{x,y,z\}} \mathcal{T}_d \right) I^n
\end{equation*}
where, for instance, the x-direction transport operator is $\mathcal{T}_x I_i^n = \frac{\Delta t}{\Delta x} (F_{i+1/2} - F_{i-1/2})$.\footnote{\revtext{For the sake of this discussion, we differentiate the spatial and spectral dimensions using indices $i$ and $j$ respectively. Flattening these indices would result in the original index $k$. Additionally, the subscript $i$ is omitted ahead for simplicity.}} The upwind or the HLL flux operator can be written in the form $F_{\pm 1/2} = (\beta_{\pm1/2} O_{x\pm} + \alpha_{\pm1/2}) I^n$
where $O_{x\pm}$ is the spatial offset operator, which offsets the zones by one in the spatial x-direction. The factors $\alpha_{\pm 1/2}$ and $\beta_{\pm 1/2}$ can be shown to be of the form $\alpha_{\pm 1/2} = a_{\pm1/2}^-(\vec{x}) \, n_x \cdot \textbf{1}_{n_x^-} + a_{\pm1/2}^+(\vec{x}) \, n_x \cdot \textbf{1}_{n_x^+}$ and $\beta_{\pm 1/2} = b_{\pm1/2}^-(\vec{x}) \, n_x \cdot \textbf{1}_{n_x^-} + b_{\pm1/2}^+(\vec{x}) \, n_x \cdot \textbf{1}_{n_x^+}$ where \textbf{1} is the indicator function, $n_x(\theta, \phi)$ is the $x$-component of the direction vector \textbf{n}, and $n_x^+ \equiv n_x \ge 0$, $n_x^- \equiv n_x < 0$. Only the spatial parts of $\alpha$ and $\beta$, e.g. $a_i$, change between the upwind and HLL fluxes. Consequently, the total transport operator for the $x$-direction is 
\begin{equation*}
\mathcal{T}_x = \frac{\Delta t}{\Delta x} \sum_{w=\pm} \left(b_{+1/2}^w O_{x+} + a_{+1/2}^w - a_{-1/2}^w - b_{-1/2}^w O_{x-}\right) n_x \cdot \textbf{1}_{n_x^w},
\end{equation*}
which clearly has a rank $(=n)$ equal to two, as the piece in parentheses is purely spatial. The Rusanov flux, on the other hand, involves a sum of angle dependent flux terms and the solution term resulting in a transport operator of the form
\begin{align}
    \mathcal{T}_x 
    = \frac{\Delta t}{\Delta x} \Big[
    \left( b_{+1/2}^1 O_{x+} + a_{+1/2}^1 - a_{-1/2}^1 - b_{-1/2}^1 O_{x-} \right) n_x\notag \\
    + \left( b_{+1/2}^2 O_{x+} + a_{+1/2}^2 - a_{-1/2}^2 - b_{-1/2}^2 O_{x-} \right)
    \Big]\notag
\end{align}
Again, the number of terms in the above sum indicates that the rank of the Rusanov operator is also equal to two. Similar arguments apply for the other dimensions for all three fluxes discussed, giving rise to the relation $n = 2d_X$.

\subsection{Speedup}
The most computationally expensive step occurs when the solution is rounded after the transport update. Rounding algorithms typically scale as $O(dNr^3)$. When dealing with a sum of TTs, an additional factor of $n$ enters the complexity. In particular, if the TT sum is formed explicitly, the cost increases by a factor of $(n+1)^3$. More efficient implementations can reduce this overhead by exploiting the block-diagonal structure of the cores in a TT sum, allowing the complexity to scale linearly with $n$. We refer the reader to \cite{al2023randomized} for more details. In this work, however, we do not analyze the additional computational cost arising from rounding the intermediate TT sum in detail, as it is already incorporated into the reported speedup estimates.

\subsection{Compression}
Explicitly forming the TT sum significantly increases intermediate storage requirements. However, both the Gram-based and randomized rounding algorithms can operate directly on a list of summands without explicitly constructing this aggregated TT. This avoids forming the full green TT and therefore reduces the peak storage required prior to rounding.

We now estimate the magnitude of this intermediate storage when pre-rounded tensors are taken into account, using the storage factor $k_s$. In our setting, $n=2$ for 1D spatial problems and $n=4$ for two-dimensional problems. The precise value of $k_s$ depends on the discretization and TT ranks, making it difficult determine exactly. Nevertheless, we can establish approximate bounds. We assume that the TT ranks exceed $n$. Although this assumption may not hold during the initial few timesteps, this is not problematic since memory constraints typically arise only at later stages of the simulation. During the intermediate transport step, the TT ranks get scaled by $n+1$ and include an additional additive term, which implies a lower bound of $n+1$ on $k_s$. The lower bound provides a good estimate when either of the boundary cores ($Z$ or $\Phi$) dominates the storage. When the SVD-based rounding algorithm is used, a rough upper bound for $k_s$ is $(n+1)^2$, since the full TT sum must be explicitly formed prior to truncation. This upper bound is approached when the central core ($\Theta$) dominates the storage. In contrast, alternative rounding algorithms that avoid forming an explicit TT sum typically yield values of $k_s$ close to $n+1$, independent of the discretization and TT ranks. In our case, we generally operate closer to the lower bound for $k_s$ because $N_\nu N_X \gg N_\theta, N_\phi$.

\begin{figure}
    \centering
    \includegraphics[width=1.0\linewidth]{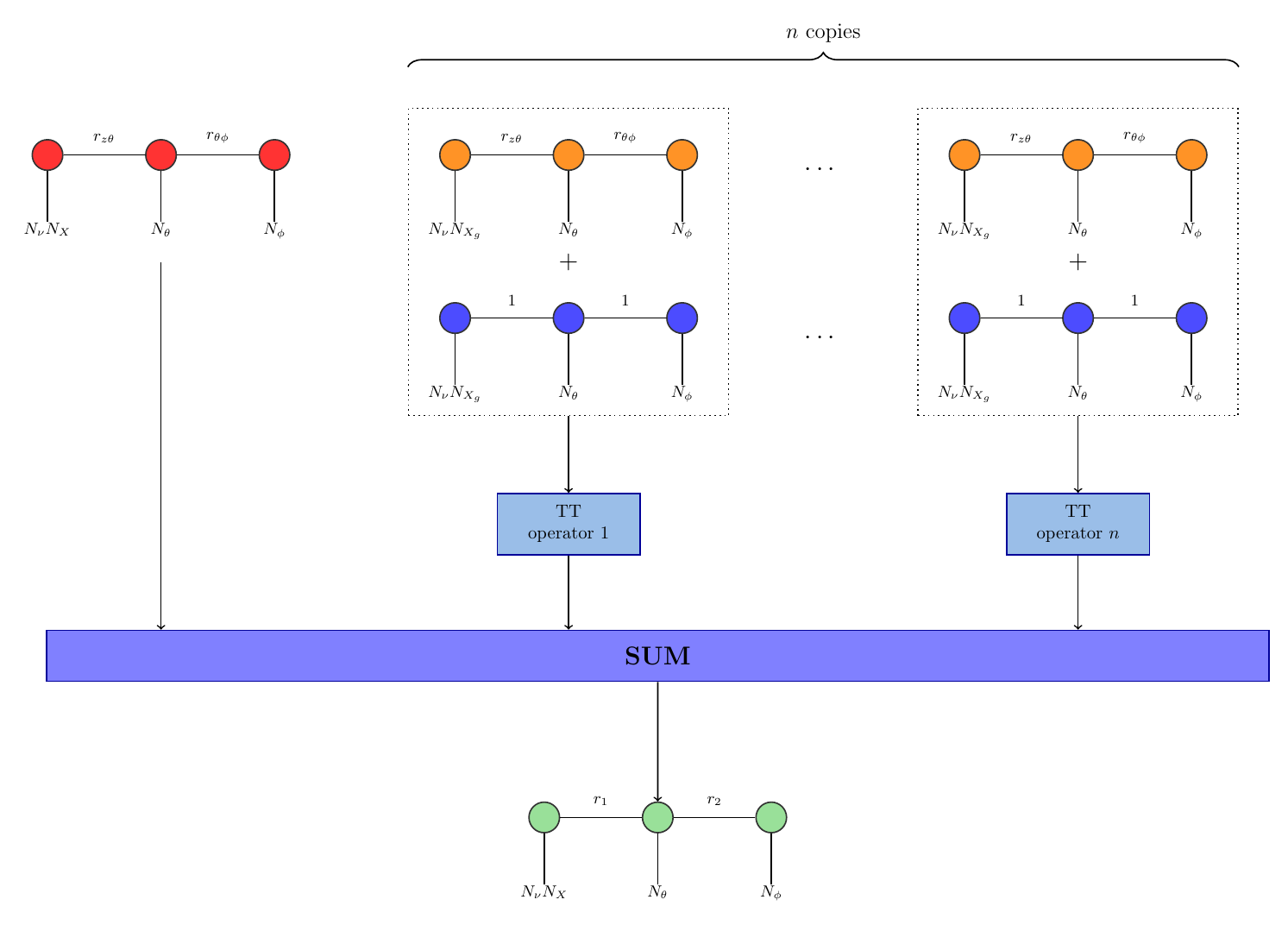}
    \caption{\revtext{Illustration of the transport update procedure. The TT with red nodes represents the solution at the beginning of the timestep. The orange TT shows the solution after padding the spatial core with ghost-cell values. The blue TT corresponds to a rank-1 Dirichlet tensor that contains boundary values in the ghost cells and zeros elsewhere. The green TT denotes the solution after the transport update. Here, $N_{X_g} = N_X + N_g$, where $N_g$ is the number of ghost cells. The transport operator is expressed as a sum of $n$ rank-1 TT operators and is applied to the sum of the ghost-cell-padded solution and the Dirichlet boundary-condition TT. In practice, this is implemented by applying each operator separately to the solution and boundary-condition tensors. This creates $n$ additional ghost-cell-padded copies of the original solution (orange). Applying operator $i$ produces an updated tensor $TT^i$ that retains the same TT ranks $\rtwo, \rthree$, while the boundary-condition TT remains rank-1. The resulting tensors are added to the solution at the beginning of the timestep to form the transport update. Forming the TT sum explicitly resutls in the green network with increased TT ranks $r_1=(n+1) \cdot \rtwo + n$ and $r_2=(n+1) \cdot \rthree + n$. This result is rounded before the source update step.}}
    \label{fig:trans_op}
\end{figure}

In Figure \ref{fig:kred}, we plot the time evolution of the factor $k_s$ for the 2D hohlraum problem (left) and the 1D Gaussian diffusion problem (right) using all 3 rounding algorithms. We observe that $k_s$ increases during the initial timesteps. Although the underlying assumptions are not strongly satisfied—for instance, $N_X N_\nu$ is comparable to $N_\phi$ for the 2D hohlraum problem and the TT ranks in the Gaussian diffusion case are only slightly greater than $n$—the value of $k_s$ eventually stabilizes relatively close to the lower bound of $n+1$ for both problems. When the SVD-based rounding algorithm is used, it reaches values of approximately $5.3$ for the diffusion problem and about $11.3$ for the hohlraum case. In contrast, the Gram rounding algorithm and randomized rounding algorithm (with a rank increase of 4) avoid forming an explicit TT sum and therefore yield more favorable values of $k_s$, remaining close to the lower bound throughout the simulation.

\begin{figure}
    \centering
    \includegraphics[width=1.0\linewidth]{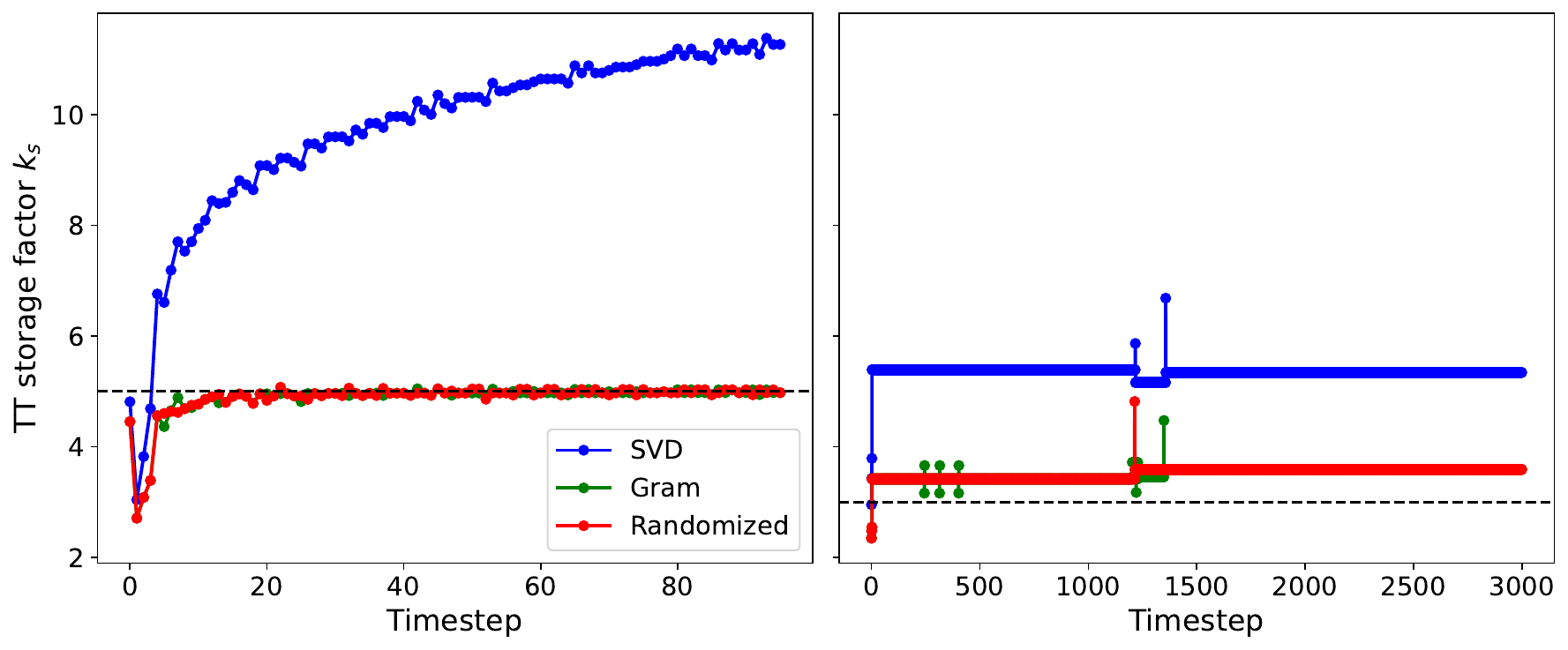}
    \caption{\revtext{Time evolution of the TT storage factor $k_s$ for the 2D hohlraum problem (left) and the 1D Gaussian diffusion problem ($\kappa_s^{max}=1100$) (right). When the SVD-based rounding algorithm is used (blue), the value of $k_s$ increases during the initial timesteps and then stabilizes relatively close to the lower bound of $n+1$ for both cases, reaching approximately $5.3$ for diffusion problem and about $11.3$ for the hohlraum problem. These values of $k_s$ represent a worst-case scenario since a TT sum must be assembled explicitly. As indicated by the red and green plots, using Gram-based or randomized rounding algorithms instead of the SVD-based approach can reduce $k_s$ and bring it much closer to the lower bound. Additionally, implementation-level optimizations such as performing rounding more frequently by forming partial sums can further limit $k_s$, potentially reducing it to values as low as $2$.}
}
    \label{fig:kred}
\end{figure}

In summary, the transport update increases the intermediate storage by a factor that depends on the spatial dimensionality of the problem, and rounding the intermediate solution becomes the most computationally intensive operation. Rounding algorithms that exploit the block-diagonal structure of a TT sum, such as Gram-based or randomized approaches, can reduce both the computational cost and the intermediate storage requirements. Lastly, we highlight that implementation-level optimizations, such as rounding more frequently through partial sum formation, can further reduce $k_s$, potentially bringing it down to values as low as $2$.

\end{revblock}

\end{document}